\documentclass[12pt]{article}

\pdfoutput=1

\usepackage[top=80pt,bottom=85pt,left=85pt,right=85pt]{geometry}
\usepackage{amssymb}
\usepackage{amsmath}
\usepackage{graphicx,color,subfigure}
\usepackage{float}
\usepackage{sectsty}
\usepackage{cite}
\usepackage[debug,pageanchor=false]{hyperref}
\definecolor{link}{rgb}{.8,.15,.1}
\hypersetup{colorlinks=true,linkcolor=link,citecolor=link,urlcolor=link,linktocpage}

\newcommand{\zz}{\mathbb{Z}}

\newcommand{\del}{\partial}

\makeatletter
\@addtoreset{equation}{section}
\makeatother

\begin{document}

\begin{titlepage}

\begin{center}

\vskip .5in 
\noindent

{\Large \bf{6d holographic anomaly match as a continuum limit}}

\bigskip\medskip

Stefano Cremonesi$^1$ and Alessandro Tomasiello$^2$\\

\bigskip\medskip
{\small 

$^1$ Department of Mathematics, King's College London, \\
The Strand, London WC2R 2LS, United Kingdom
\\	
\vspace{.3cm}
$^2$ Dipartimento di Fisica, Universit\`a di Milano--Bicocca, \\ Piazza della Scienza 3, I-20126 Milano, Italy \\ and \\ INFN, sezione di Milano--Bicocca
	
}

\vskip .5cm 
{\small \tt stefano.cremonesi@kcl.ac.uk, alessandro.tomasiello@unimib.it}
\vskip .9cm 
     	{\bf Abstract }
\vskip .1in
\end{center}

\noindent
An infinite class of analytic AdS$_7\times S^3$ solutions has recently been found. The $S^3$ is distorted into a ``crescent roll'' shape by the presence of D8-branes. These solutions are conjectured to be dual to a class of ``linear quivers'', with a large number of gauge groups coupled to (bi-)fundamental matter and tensor fields. In this paper we perform a precise quantitative check of this correspondence, showing that the $a$ Weyl anomalies computed in field theory and gravity agree. In the holographic limit, where the number of gauge groups is large, the field theory result is a quadratic form in the gauge group ranks involving the inverse of the $A_N$ Cartan matrix $C$. The agreement can be understood as a continuum limit, using the fact that $C$ is a lattice analogue of a second derivative. The discrete data of the field theory, summarized by two partitions, become in this limit the continuous functions in the geometry. Conversely, the geometry of the internal space gets discretized at the quantum level to the discrete data of the two partitions.

\noindent

\vfill
\eject

\end{titlepage}

\tableofcontents


\section{Introduction} 
\label{sec:intro}

In dimensions higher than four, a Yang--Mills theory becomes strongly coupled at high energies: this signals non-renormalizability and often means the theory is not sensible, much like for Einstein's gravity in dimensions higher than two. String theory constructions provide several examples where a gauge theory is ``UV-completed'' by a CFT: namely, there exists a CFT which flows at low energies to the gauge theory. 

A notable supersymmetric example in six dimensions is the class of so-called ``linear quiver'' theories, where one has a chain of gauge groups, coupled to (bi)fundamental hypermultiplets and to tensor multiplets. These can be engineered in string theory by placing D-branes on orbifold singularities \cite{intriligator-6d,intriligator-6d-II} or more generally with an NS5--D6--D8-brane system \cite{brunner-karch,hanany-zaffaroni-6d}. In these theories, the inverse squared Yang--Mills couplings are promoted to scalar fields: there is a point in the moduli space of vacua where all of them vanish, and the theory is strongly coupled.  The string theory engineering suggests that this point should actually be a CFT.

This picture was recently strengthened by holography. A classification of type II AdS$_7$ solutions was given in \cite{afrt}; in massive IIA an infinite series of solutions was found. These solutions were conjectured to be dual to the CFTs described above in \cite{gaiotto-t-6d}.%
\footnote{AdS solutions dual to 
linear quiver 
SCFTs in four and three dimensions were described in \cite{gaiotto-maldacena,assel-bachas-estes-gomis}.}
 Later, their analytical expression was found \cite{10letter}. The internal space $M_3$ is an $S^2$-fibration over an interval, so that the topology is that of an $S^3$; the geometry is back-reacted upon by D8-branes. A sketch of the internal geometry evokes the shape of a ``crescent roll'';%
\footnote{The first to suggest this metaphor was probably X.~Yin.} see figure \ref{fig:M3-D8}. Up to orbifolds and orientifolds, these are the most general AdS$_7$ solutions in perturbative type II. 
(Further generalizations can be engineered in F-theory \cite{heckman-morrison-vafa,dhtv,heckman-morrison-rudelius-vafa}.) 

In this paper, we are going to give strong evidence for the conjectural identification of \cite{gaiotto-t-6d} between the linear quiver CFTs and the crescent roll solutions. The evidence consists of a systematic comparison of the so-called $a$ anomaly on both sides. This is the part of the Weyl anomaly which is proportional to the Euler density; it is generally thought to be a measure of the number of ``degrees of freedom'' of a field theory. For example, it has been shown never to increase in RG flows in two \cite{zamolodchikov} and four \cite{komargodski-schwimmer} dimensions; for a theory with a holographic dual, this property can be argued in general \cite{girardello-petrini-porrati-zaffaroni,freedman-gubser-pilch-warner}. 

On the field theory side, we computed the $a$ anomaly using the Lagrangian formulation away from the CFT point in the moduli space, where conformal invariance has been spontaneously broken and the Yang--Mills couplings are finite. One can use the relation \cite{cordova-dumitrescu-intriligator-a6} of $a$ to the anomalies of R-symmetry and diffeomorphisms, which are not broken and can be reliably computed away from the CFT point. While the number of fields presumably decreases a lot in the RG flow from the CFT to the Lagrangian theory, some of the remaining fields obtain non-trivial gauge transformations that make up for the loss. In this case, this is a Green--Schwarz--West--Sagnotti (GSWS) \cite{green-schwarz-west,sagnotti} mechanism; its precise contribution can be determined by imposing cancelation of gauge anomalies. This method was used in \cite{intriligator-a6,osty-a6} to compute anomalies for a vast class of six-dimensional theories; here we apply it to the most general linear quiver, and extract the term that dominates in the holographic limit. This turns out to involve, in this case, taking to infinity \emph{the number $N-1$ of gauge groups}, rather than each of the individual ranks. If the gauge groups are SU$(r_i)$, $i=1,\ldots,N-1$, in this limit we obtain
\begin{equation}\label{eq:a-intro}
	a = \frac{192}7 \sum C^{-1}_{ij}r_i r_j\ ,
\end{equation}
where $C$ is the Cartan matrix for $A_{N-1}$. 

On the gravity side, $a$ is computed as the volume of the internal space $M_3$ in Einstein frame, normalized in a certain way to the AdS$_7$ radius. This particular combination actually appears in other holographic estimates of the number of degrees of freedom, at leading order. For example in four dimensions $a$ and $c$ happen to coincide \cite{henningson-skenderis} up to string-theory corrections. Similarly, for the six-dimensional $\mathcal{N}=(1,0)$ theories studied in this paper, it turns out that up to string-theory corrections 
the coefficients $c_i$ of the three independent Weyl-invariants are all proportional to $a$. The reason is that $a$ and $c_i$ are all linear in the four coefficients of the anomalies of the R-symmetry and diffeomorphisms \cite{cordova-dumitrescu-intriligator-a6,beccaria-tseytlin-ci}, and only one of these anomaly coefficients determines the leading behavior in the holographic limit.
Also, the same combination appears in the thermal free energy coefficient ${\cal F}_0$, which appears in ${\cal F}\sim {\cal F}_0 T^d {\rm Vol}$.

A computation of this coefficient was performed in \cite{10letter} for a couple of examples. For instance, for a symmetric solution with two D8-branes, the result in \cite{10letter} is the complicated-looking\footnote{Here and in the following we will set to 1 the anomaly of an abelian $(2,0)$ tensor, as in \cite{cordova-dumitrescu-intriligator-a6}.}
\begin{equation}\label{eq:a-ex}
	a_{\rm hol}=\frac{16}7 k^2\left(N^3-4 N k^2 + \frac{16}5 k^3\right)\ 
\end{equation}
where $k$ is another integer of order $N$ characterizing the quiver (see figure \ref{fig:2D8} below). This exhibits the $N^3$ scaling typical of fivebranes \cite{klebanov-tseytlin-n3}. Notice, however, that $k\sim N$: hence this should be thought of as a polynomial of overall degree 3 in $N$ and $k$; all the terms come from supergravity, not from string-theory corrections, which we do not consider in this paper. 
Applying the field theory result (\ref{eq:a-intro}) to this case, one gets exactly (\ref{eq:a-ex}), matching all the coefficients.

Encouraged by this result, we have performed this holographic computation in general, obtaining a perfect match with the field theory result. Although the detailed comparison is complicated, we can already sketch a heuristic argument here. The gravity solutions depend on a certain function $q(z)$, which in appropriate coordinates is piecewise linear. This function actually interpolates the discrete graph of (half of) the gauge ranks $r_i$ (see figure \ref{fig:r-plot}). The holographic computation $a_{\rm hol}$ reduces to an integral of $q$ times a second primitive of $q$; schematically, $a_{\rm hol}\propto \int q \frac1{\del^2} q$. But the Cartan matrix $C$ of $A_{N-1}$ can be viewed as (minus) a discrete second derivative, as is evident from writing it as $(C r)_i = -r_{i+1}+ 2 r_i - r_{i-1}$. Since the holographic limit involves taking $N\to \infty$, we can think of it as some kind of continuum limit, and 
\begin{equation}\label{eq:hol-comp}
	a =\frac{192}7 \sum C^{-1}_{ij}r_i r_j \ \buildrel \text{hol. limit} \over \longrightarrow \ 
	 a_{\rm hol}= \frac{192}7 \int 4q(z) \frac1{\del_z^2} q(z) dz\ .
\end{equation}
While this argument might feel a little schematic, we make the continuum limit more precise and present the calculation in full detail below, and we indeed obtain full agreement between the field theory and gravity computations. 

Turning the result on its head, we can say that at finite $N$ the field theory gives some kind of quantum discretization of the gravity solution, where the function $q$ entering the metric gets discretized by the graph of the $r_i$. It is of course often emphasized in holography that the field theory side provides a quantum definition of the corresponding gravity solution, but this class of examples gives a particularly clear example of this.

The paper is organized as follows. In section \ref{sec:rev} we review the linear quiver six-dimensional field theories, and the AdS$_7$ solutions conjectured in \cite{gaiotto-t-6d} to be their gravity duals. In section \ref{sec:an} we perform the computation of $a$ in field theory, and extract the term that dominates in the holographic limit. In section \ref{sec:hol} we compute $a_{\rm hol}$ and we compare it with $a$, making (\ref{eq:hol-comp}) more precise.



\section{6d linear quivers and their holographic duals} 
\label{sec:rev}

In this section, we will review the six-dimensional linear quiver $(1,0)$ theories of \cite{brunner-karch,hanany-zaffaroni-6d} and their gravity duals, proposed in \cite{gaiotto-t-6d} to be the AdS$_7$ solutions of \cite{afrt,10letter}. We will also work out in full generality certain details of the gravity solutions, such as the explicit positions of the D8-branes, which in \cite{afrt,10letter} were only computed in some examples.

\subsection{The field theories} 
\label{sub:ft}

The theories were originally inferred to exist from brane configurations involving NS5-branes, D6-branes and D8-branes. The NS5-branes are extended along directions $0,\ldots,5$; the D6-branes along $0,\ldots,6$; the D8-branes along all directions except $6$. See figures \ref{fig:D8-in}, \ref{fig:D8-out}, which we will explain in detail later, for an example. (Both brane configurations engineer the same theory: they are related by Hanany--Witten moves \cite{hanany-witten}.) When the NS5-branes are not on top of each other, the system is described by a field theory that can be read off \cite{hanany-zaffaroni-6d,brunner-karch} using the strategy originally outlined in \cite{hanany-witten} for three-dimensional field theories. When the NS5-branes are on top of each other, we lose a Lagrangian description and we expect interesting phenomena. 

If $N$ is the number of NS5-branes, the quivers consist of $N-1$ vector multiplets $(A_{\mu i},\lambda_{i \alpha},D_i)$ with gauge groups U$(r_i)$, $i=1,\ldots, N-1$; hypermultiplets $(h_i,\psi_{i \dot \alpha})$, $i=1,\ldots,N-2$, in the bifundamental $\overline{\mathbf{r_i}}\otimes \mathbf{r_{i+1}}$, and $f_i$ hypermultiplets $(\tilde h^{a_i}_i,\tilde\psi^{a_i}_{i \dot \alpha})$, $i=1,\ldots,N-1$, in the fundamentals $\mathbf{r_i}$; tensor multiplets $(\Phi_i, \chi_{i \alpha},B_{i \mu\nu})$, $i=1,\ldots, N$, where the two-form potentials $B_{i\mu\nu}$ have self-dual field-strengths $H_{i\mu\nu\rho}$; and, finally, linear multiplets $((\pi_i, C_i), \xi_{i \dot\alpha})$, $i=1,\ldots, N$, where $\pi_i$ are SU(2)$_R$ triplets of noncompact scalars while $C_i$ are SU(2)$_R$ singlet periodic scalars, see for example \cite{douglas-moore,park-taylor}. The real scalars $\Phi_i$ in the tensor multiplets enter the kinetic terms of the gauge groups according to $(\Phi_{i+1} - \Phi_{i}){\rm Tr}|F_i|^2$; this dictates an ordering $\Phi_i < \Phi_{i+1}$, and moreover, when all the $\Phi_i$ coincide $(\Phi_i=\Phi_{i+1} \forall i)$ the effective gauge couplings of all gauge groups are divergent, the theory becomes strongly coupled and contains tensionless strings. In fact the $\Phi_i$ realize the positions of the NS5-branes along $x^6$, and the strong coupling point we just mentioned corresponds to the NS5-branes being on top of each other. When the scalars $\Phi_i$ in the tensor multiplets take different expectation values, the theory is said to be on the tensor branch. 
Similarly, the triplets $\pi_i$ realize the positions of the NS5-branes along $x^{7,8,9}$; $C_i$ may be thought of as the positions along $x^{10}$ if there is an M-theory uplift. (From the four scalars in each hypermultiplet one can form hyper-momentum maps for the U(1)$_i$ centers of the U$(r_i)$ gauge groups, which are equated to $\pi_i$ by the equations of motion.)

Given all these ingredients, at generic points on the tensor branch where $\Phi_i \neq \Phi_{i+1}$ one can write the equations of motion of these theories (or equivalently a ``pseudo-action'' on top of which one has to impose the self-duality constraints $H_i= *H_i$ by hand). This can be done for example by specializing the ``tensor hierarchy'' actions \cite{samtleben-sezgin-wimmer,samtleben-sezgin-wimmer-hyper}, setting to zero their St\"uckelberg-like terms $h^t_I$ and $g^{Js}$, but keeping their $d^I_{rs}$. (Further work on these theories has also produced Lagrangians whose equations of motion also contain the self-duality constraints \cite{bandos-samtleben-sorokin}.)

There is a further subtlety: the U$(1)$ subgroup in each of the U$(r_i)$ gauge groups actually suffers from a further anomaly, which is presumed to be canceled \cite{Berkooz:1996iz,hanany-zaffaroni-6d,park-taylor} by a GSWS mechanism involving this time an anomalous transformation of the periodic scalars $C_i$ in the linear multiplets, which gives a mass to the U$(1)$ factors in the gauge groups via a St\"uckelberg mechanism. This effect was not included in \cite{samtleben-sezgin-wimmer,samtleben-sezgin-wimmer-hyper} and its stringy origin has not been worked out in detail.%
\footnote{We thank T.~Dumitrescu for interesting discussions about this point.} Since we are interested in the low energy physics and in computing anomalies, we will proceed by forgetting the massive U(1)'s, and considering SU$(k_i)$ gauge groups.

\begin{figure}[ht]
	\centering
		\includegraphics[width=6cm]{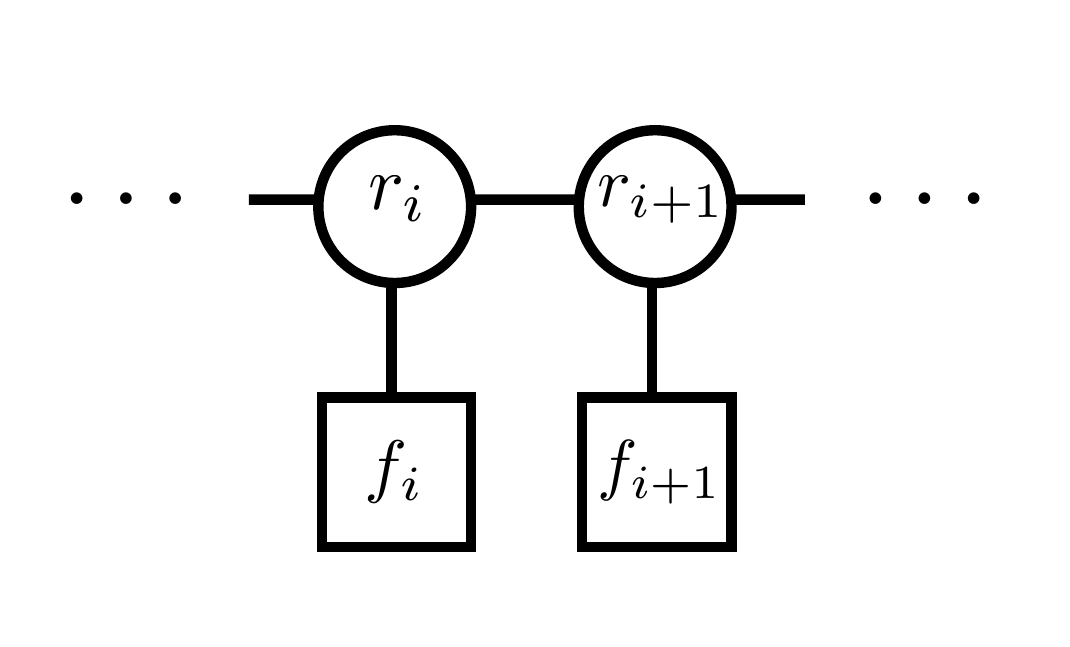}
	\caption{\small The general structure of a linear quiver.}
	\label{fig:r-f}
\end{figure}

\begin{figure}[p]
\centering	
	\subfigure[\label{fig:quiver}]{\includegraphics[width=14cm]{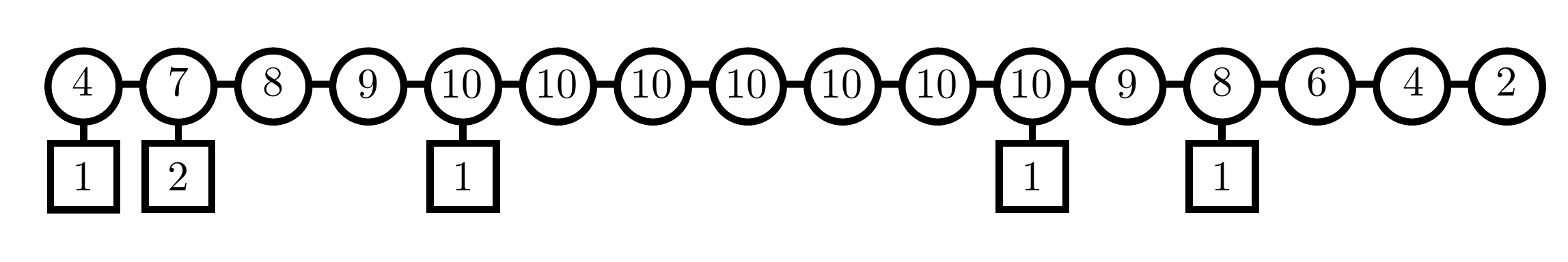}}\\
	\subfigure[\label{fig:r-plot}]{\includegraphics[width=14cm]{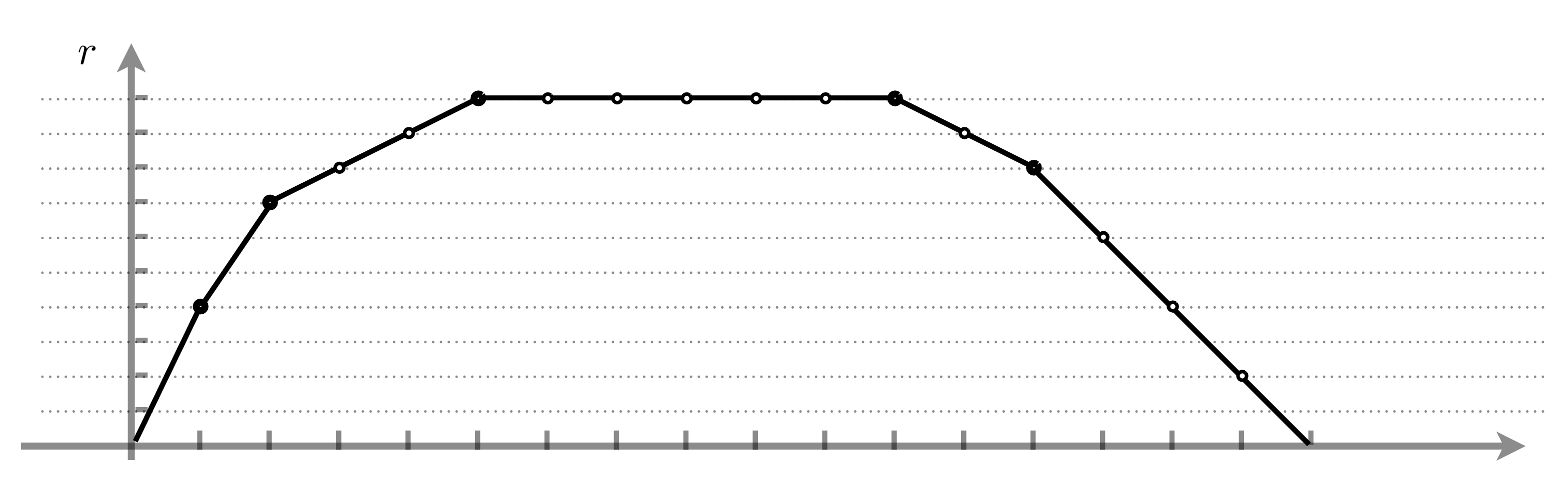}}\\
	\subfigure[\label{fig:s-plot}]{\includegraphics[width=14cm]{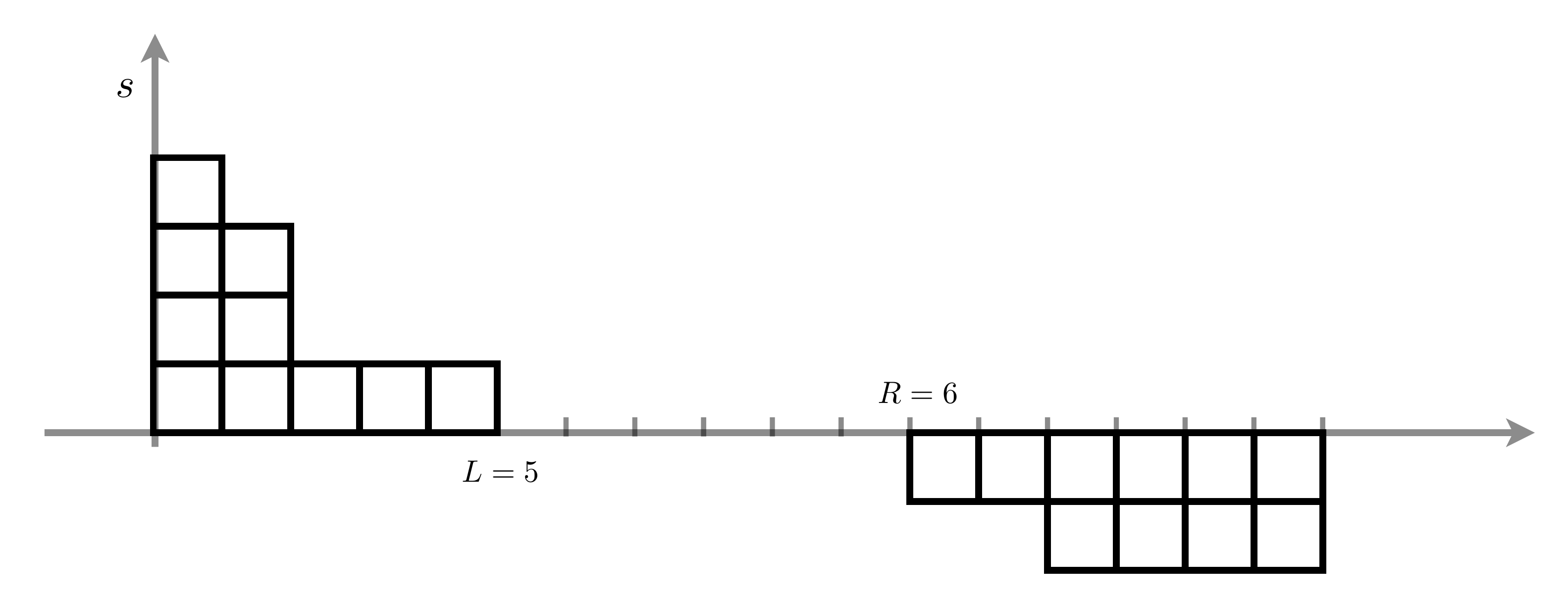}}\\
	\caption{\small In \subref{fig:quiver}, an example of linear quiver. As described in the text, round nodes represent gauge groups, square nodes flavor symmetries. Links represent hypermultiplets; horizontal links also have tensor multiplets associated to them. In \subref{fig:r-plot} we plot the numbers of colors $r_i$, 
as a function of the position $i$ in the quiver. We added a linear interpolation to guide the eye. The bigger dots indicate points where the slope changes; these are the positions where flavors are present, and the change in slope equals the number of flavors. In \subref{fig:s-plot} we plot the $s_i=r_i-r_{i-1}$; this can be thought of as the derivative of the linear interpolation in \subref{fig:r-plot}. We have filled in the plot with boxes, that define two Young diagrams $\rho_{\rm L}$, $\rho_{\rm R}$.}
	\label{fig:cft}
\end{figure}

\begin{figure}[p]
\centering	
\subfigure[\label{fig:D8-in}]{\includegraphics[width=14cm]{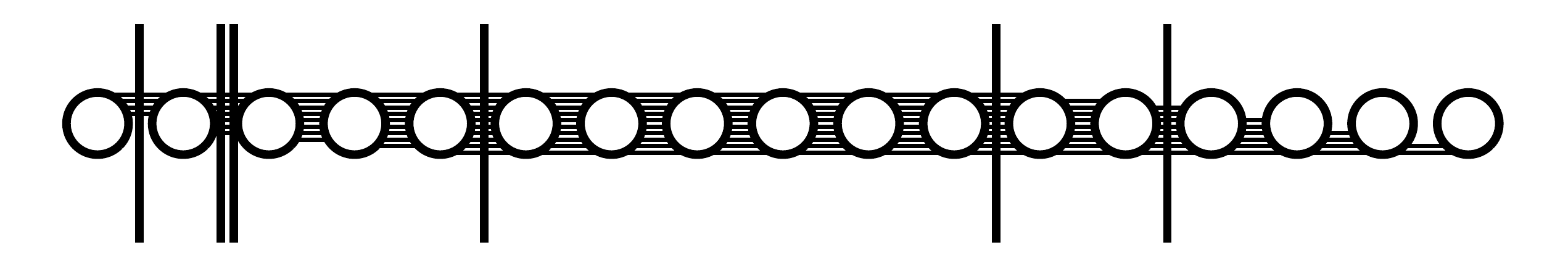}}\\
\subfigure[\label{fig:D8-out}]{\includegraphics[width=10cm]{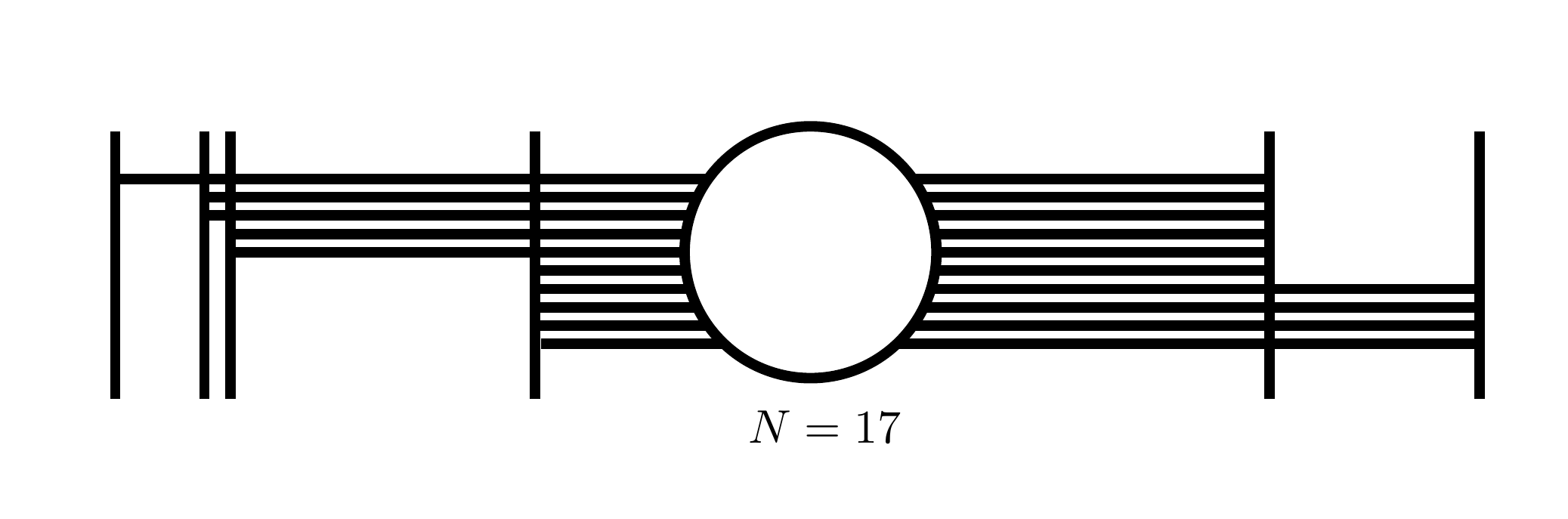}}\\
\subfigure[\label{fig:M3-D8}]{\includegraphics[width=6.5cm]{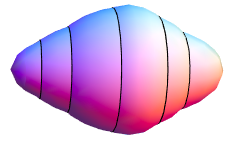}}
	\caption{\small In \subref{fig:D8-in} and \subref{fig:D8-out} we see two versions of the brane system that engineers the particular quiver in figure \ref{fig:quiver}, related by Hanany--Witten moves. In both cases, round dots represent NS5-branes; horizontal lines represent D6's; vertical lines represent D8's. In \subref{fig:D8-in} we see the system in a configuration where the quiver can be read off easily: the segment between the $i$-th and $(i+1)$-th NS5-branes contains $r_i$ D6-branes, and $f_i$ D8-branes intersecting them. The two Young diagrams can be read off intuitively from both pictures \subref{fig:D8-in} and \subref{fig:D8-out}. Focusing for example on $\rho_{\rm L}$, in \subref{fig:D8-in} we see that there are 1 D8 in the first segment, 2 in the second, 1 in the fifth: these represent the drops $s_i-s_{i+1}=f_i$ in the Young diagram. In \subref{fig:D8-out} we see even more directly that there are  1 D8 with $\mu=1$ D6-branes ending on it, 2 D8's with $\mu=2$ D6's ending on them, 0 D8's with $\mu=3$, 0 D8's with $\mu=4$, 1 D8's with $\mu=5$; these are the $f_i= s_i-s_{i+1}$ associated to $\rho_{\rm L}$.
Finally in \subref{fig:M3-D8} we see an artist's impression of the shape of the internal $M_3$ in the AdS$_7$ solution. The D8's are represented by the black lines. There are as many D8-brane stacks as in the brane pictures (the two D8's with $\mu=2$ are on top of each other). These D8 stacks are in correspondence with the flavors in figure \ref{fig:quiver}.}
	\label{fig:branes}
\end{figure}

The field theory on the tensor branch can be now summarized as a quiver. Each round node with number $r$ represents an SU$(r)$ gauge group; each link between two round nodes corresponds to a bifundamental hypermultiplet, and a tensor multiplet (two more tensor multiplets are associated to the extremal NS5-branes corresponding to $I=1,N$); and finally, links from a round node to a square node with number $f$  represents $f$ hypermultiplets in the fundamental representation of the gauge group (and antifundamental representation of a U$(f)$ flavor symmetry). See figure \ref{fig:r-f}, and figure \ref{fig:quiver} for a particular example.

For an SU$(r)$ vector coupled to $f$ flavors in the fundamental or antifundamental, gauge anomaly cancelation dictates $f=2r$. (We will rederive this constraint in section \ref{sec:an}.)
For our quiver, this implies 
\begin{equation}\label{eq:rifi}
	2r_i-r_{i+1}-r_{i-1}=f_i\ .
\end{equation}
Intuitively, this says that the numbers of flavors $f_i$ are a sort of minus ``discrete second derivative'' of the numbers of colors $r_i$. 
As in lattice QFT, one can also introduce forward and backward discrete derivatives $(\del r)_i \equiv r_{i+1}-r_i$, $(\del^* r)_i \equiv r_{i}-r_{i-1}$, so that $f=-\del \del^* r$. 
Since the $f_i$ are by definition non-negative, it follows that the function $r_i$ is concave. Thus, it will increase from zero ($r_0\equiv 0$), possibly have a plateau in the middle, and then decrease to zero again ($r_N\equiv 0$). See figure \ref{fig:r-plot}.

It is also convenient to introduce the ``slopes'' 
\begin{equation}\label{eq:si}
	s_i = r_i-r_{i-1} = (\del^* r)_i\ ,
\end{equation}
in terms of which 
\begin{equation}\label{eq:fi}
	f_i= s_i - s_{i+1} = - (\del s)_i\ .
\end{equation}
From what we just said, it follows that the slopes $s_i$ define a decreasing 
function. Its plot defines visually two Young diagrams, made from the positive $s_i$ on the left and the negative $s_i$ on the right, possibly separated by a zero region (which corresponds to the plateau we mentioned earlier). See figure \ref{fig:s-plot}. These two Young diagrams $\rho_{\rm L}$ and $\rho_{\rm R}$ provide a convenient way of parameterizing the theories we are considering, in the sense that the data of the ranks $r_i$ and $f_i$ can be completely reconstructed from them and from the number $N$.%
\footnote{In the brane construction of the linear quivers, the Young diagrams $\rho_{\rm L}$ and $\rho_{\rm R}$ encode the boundary conditions of a stack of $k$ D6-branes ending on two stacks of D8-branes \cite{gaiotto-t-6d}, in the spirit of \cite{gaiotto-witten-1}.}
In other words, the CFT$_6$'s we are considering in this paper can be parameterized as 
\begin{equation}
	{\cal T}^N_{\rho_{\rm L}, \rho_{\rm R}}\ .
\end{equation}
By construction, the two Young diagrams have the same number of boxes. Indeed, let us call $L$ the depth of the left Young diagram, $R$ the depth of the right one, and $k$ the maximum rank, that is the height of the plateau when it is present. From (\ref{eq:si}) we see that  
\begin{equation}
	k = \sum_{i=1}^L s_i = -\sum_{j=N-1}^R s_i \ .
\end{equation}
The Young diagrams can also be read off easily from the brane configurations: see again figures \ref{fig:D8-in}, \ref{fig:D8-out}, and cf.~figure \ref{fig:cft}.

\subsection{The gravity duals} 
\label{sub:ads}

We will now describe the AdS backgrounds in massive type IIA supergravity which have been proposed \cite{gaiotto-t-6d} as gravity duals to the field theories we just described. Originally the problem of finding AdS$_7\times M_3$ solutions in type II supergravity was reduced to a certain ODE system in \cite{afrt}, where some solutions were found numerically. More recently their analytical form was found \cite{10letter}. The metrics have a certain local expression that depends on a single parameter; one can then glue several ``pieces'' of this local metric along D8-branes. This gluing was illustrated in \cite{10letter} in a couple of examples; here we will also complete the exercise of working this gluing out along an arbitrary number of D8's. This will be needed for the holographic computation of the anomaly in section \ref{sec:hol}. 

\subsubsection{Solutions} 
\label{ssub:sol}

The metric in string frame reads 
\begin{equation}\label{eq:ds2}
	ds^2_{10}=e^{2A}\left( ds^2_{{\rm AdS}_7} -\frac1{16}\frac{\beta' dy^2}{\beta y}+ \frac{\beta/4}{4\beta- y \beta'} ds^2_{S^2}\right)\ ,\qquad e^{2A}\equiv\frac49\sqrt{-\frac{\beta'}y} 
\end{equation}
and the dilaton is 
\begin{equation}\label{eq:ephi}
	e^\phi= \frac{(-\beta'/y)^{5/4}}{12\sqrt{4 \beta- y \beta'}} \ .
\end{equation}
Here $\beta$ is a function of $y$ such that $q\equiv -4 y \frac{\sqrt{\beta}}{\beta'}$ obeys
\begin{equation}\label{eq:dq2}
	\del_y(q^2) = \frac29 F_0 \ ,
\end{equation}
with $F_0$ the Romans mass. There are also the fluxes 
\begin{equation}\label{eq:F2H}
\begin{split}
	&F_2 = y\frac{\sqrt{\beta}}{\beta'}\left(4+\frac{F_0}{18y}\frac{(\beta')^2}{4 \beta - y \beta'} \right){\rm vol}_{S^2} \ ,\\
	&H= -9\left(-\frac y{\beta'}\right)^{1/4}\left(1-\frac{F_0}{108y}\frac{(\beta')^2}{4 \beta-y \beta'} \right){\rm vol}_{M_3}\ .
\end{split}
\end{equation}

The simplest solution has $F_0=0$. From (\ref{eq:dq2}) we get
\begin{equation}\label{eq:beta-ml}
	\sqrt\beta=\frac2k(R_0^2 - y^2) \qquad (F_0=0)\ .
\end{equation}
This is a reduction of the AdS$_7\times S^4/\zz_k$ solution of eleven-dimensional supergravity (see \cite[Sec. 5.1]{afrt} for a discussion in slightly different coordinates). It has $k$ D6-branes at the north pole $y=-R_0$ and $k$ anti-D6-branes at the south pole $y=R_0$. 

It is more interesting to consider solutions with $F_0\neq 0$. From (\ref{eq:dq2}), we see that $q^2$ must be a linear function $\frac29 F_0(y-y_0)$, and thus we find \cite{10letter}
\begin{equation}\label{eq:beta-s}
	\sqrt{\beta}= -2\int \frac{y dy}{\sqrt{\frac29 F_0 (y-y_0)}}= \sqrt{\frac 8{F_0}}\sqrt{y-y_0}(-2y_0-y)+ \sqrt{\beta_0}\ .
\end{equation}
(We have assumed here $y_0<0$, $F_0>0$, which will be convenient later.) The easiest case is when $\beta_0=0$. Under this assumption, plugging $\beta$ in (\ref{eq:ds2}), we find that the $S^2$ shrinks at $y=y_0$ and $y=-2y_0$, so that the internal space is topologically an $S^3$. At $y=y_0$, the $S^2$ shrinks in a regular way; at $y=-2y_0$, there is a singularity, which can fortunately be interpreted physically as due to a stack of anti-D6-branes. 

If one varies the integration constant $\beta_0$ in (\ref{eq:beta-s}), one can obtain more general solutions with a variety of sources \cite[Sec.~5.6]{afpt}. In this paper, however, we will be more interested in another type of generalization, namely introducing D8-branes. These have the effect of changing $F_0$ as they are crossed; thus $q^2$ is no longer linear, but only piecewise linear in $y$. The effect on $\beta$ is that, in each region between two D8-branes, one gets an expression of the type (\ref{eq:beta-s}), but with different values for the integration constants $y_0$ and $\beta_0$ (as well as $F_0$, as we just mentioned). The exception is a possible region where $F_0=0$, where (\ref{eq:beta-ml}) should be used. 

We will also switch on D6-brane charge on the D8's, by having a non-trivial gauge bundle on the internal $S^2$ that they are wrapping. We will call this integer charge $\mu$. To completely determine the solution, we should 
know where the D8-branes are located. This is 
fixed by supersymmetry, by the formula
\begin{equation}\label{eq:qD8}
	q|_{\rm D8}= \frac12(-n_2 + \mu n_0)\ ,
\end{equation}
where 
\begin{equation}\label{eq:flux-int}
	n_0 \equiv 2\pi F_0 \ ,\qquad n_2 = \frac1{2\pi}\int_{S^2} (F_2-B F_0)  
\end{equation}
are the flux integers. They both jump across the D8, but (\ref{eq:qD8}) remains invariant. 

(\ref{eq:qD8}) comes about in several related fashions. Supersymmetry fixes the fluxes as in (\ref{eq:F2H}). From these one can obtain a local formula for the $B$ field; imposing its continuity across a D8 leads to (\ref{eq:qD8}). One finds (\ref{eq:qD8}) again by imposing the Bianchi identity for $F_2$, with the correct source terms. Finally, one also recovers (\ref{eq:qD8}) with a probe calculation using calibrations. For more details, see \cite[Sec.~4.8]{afrt}.

Note that $\mu$ and $n_2$ are not invariant under large gauge transformations, but (\ref{eq:qD8}) is. For definiteness, in the remainder of the paper we make the following gauge choice. The $B$-field potential is chosen to vanish at the North and South poles of $M_3$. Since its flux through $M_3$ is $N$, this requires that we make a total of $N$ units of large gauge transformations between the poles.  
To keep $n_2$ invariant, we perform these large gauge transformations in the massless region. We will therefore distinguish between D8-branes in the region to the North and D8-branes in the region to the South of the massless region where large gauge transformations are performed. 
In the NS5--D6--D8 brane configuration, our gauge choice corresponds to keeping all the $N$ NS5-branes together in the massless region, partitioning the D8-branes in two subsets, to the left and to the right of the NS5-branes, as depicted in figure \ref{fig:D8-out}. (Different choices are related by Hanany-Witten moves \cite{hanany-witten}, which lead to the creation of D6-branes.)

Let us now state the identification proposed in \cite{gaiotto-t-6d} between these solutions and the quivers of section \ref{sub:ft}. A quiver characterized by a sequence of $N-1$ SU$(r_i)$ gauge groups with U$(f_i)$ flavor groups attached is dual to an AdS$_7$ solution of the type discussed in this subsection, with $N=-\frac1{4\pi^2}\int H$, and 
\begin{equation}\label{eq:fD8}
	f_i \ \text{D8-branes\ of\ D6-charge}\ \mu=\begin{cases}
i & \mathrm{(North)}\\
i-N & \mathrm{(South)}
\end{cases}
\end{equation}
so that $\mu$ is positive (negative) for D8-branes in the region North (South) of the massless region where the large gauge transformations are performed.%
\footnote{If the large gauge transformations were performed south of all D8-branes --- or equivalently all NS5-branes were to the right of the D8-branes in the brane diagram, as in figure 7 of \cite{dhtv} --- there would instead be $f_i$ D8-branes of D6-charge $\mu=i$ for all $i$.}
The $k$ in (\ref{eq:beta-ml}) turns out to be the same as the $k$ we defined in field theory, namely the maximum rank.%
\footnote{\label{foot:s1}In the limit case $N-L-R=0$, there is no such massless region; these are the cases discussed in \cite[Sec. 4.2]{gaiotto-t-6d}. In such a case, we can alternatively characterize the gravity solution as having $F_0=s_1$ near the ``North Pole'' $y=y_0$. The wisdom of this choice will be apparent soon.} This correspondence was originally motivated by the similarity of the data of the brane diagrams and of the AdS$_7$ solutions (see figure \ref{fig:branes}). In the language of brane diagrams, the correspondence also says that a D8-brane on which $\mu$ D6-branes end (in the configuration where all the D8's are on the outside, as in figure \ref{fig:D8-out}) becomes in the AdS$_7$ solution a D8 with D6-charge $\mu$.


\subsubsection{D8-branes} 
\label{ssub:d8}

Let us now work out the details of such a solution. First, let us spell out what (\ref{eq:qD8}) means in terms of the quiver data. A point of notation: we will consider ``the $i$-th D8 stack'' to be the one which contains D8-branes with D6-charge $\mu=i$ or $\mu=i-N$, depending on the region; as we just saw in (\ref{eq:fD8}), this stack consists of $f_i$ D8's. We will keep saying this even if some $f_i$ might be zero. (For example, in the example of figure \ref{fig:cft} and \ref{fig:branes}, we first have a stack of $f_1=1$ D8's, then a stack of $f_2=2$ D8's; then it might be more intuitive to say that the third stack is the third non-trivial stack, consisting of one D8 with charge $\mu=5$, but we will say instead that this is the fifth stack, while the third and fourth stacks will be ``empty'' stacks with $f_3=0$ and $f_4=0$ D8-branes in them.) This slight abuse of notation will be convenient.

We can now compute easily the flux integer $n_{0,i}$ (the D8-brane charge) between the $(i-1)$-th and the $i$-th stack. 
 Thinking about the generic case where there is a region with $F_0=0$, we can start from there and go towards the North Pole $y=y_0$: to get there we have crossed $f_L + f_{L-1}+\ldots +f_1 = s_1$ D8's, so the value of the flux integer $n_0$ there is $s_1$. (This now explains footnote \ref{foot:s1}.) Going backwards towards $F_0=0$, we cross the second stack with $f_1$ D8-branes, and the flux integer $n_0$ now is $f_L + f_{L-1}+\ldots +f_2 = s_2$. In general we find 
\begin{equation}\label{eq:n0si}
	n_{0,i}=s_i\ .
\end{equation}
Along similar lines we find
\begin{equation}
	n_{2,i}=-\sum_{j=1}^{i-1} j f_j \ .
\end{equation}
This can be checked visually in figure \ref{fig:D8-out}, if we recall that in such a diagram $\mu$ is the number of D6's ending on the given D8. (For example, on the left we have first a region without D6's; then after the first stack a region with only one D6; then after the second stack a region with 5 D6's; and finally the central region with 10 D6's. Looking back at $\rho_{\rm L}$ in figure \ref{fig:s-plot}, we have $f_1=1$, $f_1+2f_2=5$, $f_1+2f_2+3f_3+4f_4+5f_5=10$.)

It is now interesting to compute the value $q_i$ of $q$ at the $i$-th stack,
applying (\ref{eq:qD8}). Given (\ref{eq:n0si}), the first value is simply $q_1=\frac{s_1}2$, which recalling (\ref{eq:si}) is also equal to $\frac{r_1}2$. More generally we have $q_i=\frac12(is_i + \sum_{j=1}^{i-1}j f_j)$. Then using (\ref{eq:fi}) $2(q_i-q_{i-1})=(i-1)(s_i-s_{i-1})+s_i +(i-1) f_{i-1}= s_i$. By induction and using (\ref{eq:si}) we have
\begin{equation}\label{eq:qiri}
	q_i = \frac12 r_i\ .
\end{equation}
Note that, according to (\ref{eq:qD8}), $2q|_{\rm D8}$ equals a D6-brane charge which is both integer quantized and invariant under large gauge transformations. (This is only possible because the D6-charge is computed on the worldvolume of D8-branes.) It was perhaps to be expected that it corresponds to the number of colors $r$ in the quiver. 

Recall now from (\ref{eq:dq2}) that $q^2$ is piecewise linear in $y$, and that its slope is $\frac29F_0$; collecting the definition (\ref{eq:flux-int}) of flux integer, (\ref{eq:n0si}), and (\ref{eq:qiri}), we have
\begin{equation}\label{eq:q2y}
	q^2(y)= \frac1{9\pi} s_{i+1} (y-y_i) + \frac14 r_i^2 \ ,\qquad y_{i}\le y \le y_{i+1}
\end{equation}
where $y_i$ is the position of the $i$-th D8 stack (and, as previously defined, $y_0$ is the position of the ``North Pole'', where the $S^2$ shrinks to zero). By evaluating this at $y=y_{i+1}$ and using (\ref{eq:si}), we also get 
\begin{equation}\label{eq:Deltay}
	\Delta y_{i+1}\equiv y_{i+1}-y_i = \frac94\pi (r_{i+1}+r_i)\ .
\end{equation}
This manipulation is actually not warranted in the massless region, where $F_0=0$ (since we divided by $s_i$). In the massless region, we have another equation:
\begin{equation}\label{eq:yRL}
	y_{\rm R}- y_{\rm L}= \frac94 k \pi (N-L-R)\ , 
\end{equation}
which is obtained using \cite[Eq.~(5.42)]{afpt} and some consequences of (\ref{eq:beta-ml}). Recall that $k\equiv \rho_{\rm L}=\rho_{\rm R}$ is the maximum rank (for example, $k=10$ in figure \ref{fig:r-plot}).

As we will see, this almost fixes the positions of all D8-branes. Before we do so, however, it proves convenient to introduce a different coordinate, which will also help a great deal in comparing the supergravity data with the field theory ones.


\subsubsection{The coordinate $z$} 
\label{ssub:z}

We have seen that the value of $q$ at the $i$-th stack is given by the $i$-th rank, (\ref{eq:qiri}). This might 
suggest some resemblance between $2q$ and the piecewise linear function that interpolates between the ranks in figure \ref{fig:r-plot}. However, this fails for two reasons. First, (\ref{eq:q2y}) shows that it is $q^2$ which is piecewise linear, not $2q$. Second, when one works out the $y_i$ values of the D8's (as we will do shortly), they are not linearly spaced.%
\footnote{One might think of using $q$ itself as a coordinate in which the D8-brane positions are linearly spaced. Unfortunately, $q$ is \emph{constant} in the massless region.} 

To fix the first discrepancy, one might simply want to define a new coordinate $z$ such that $2dq=n_0 dz$ --- so that $2q$ will be piecewise linear, with slope given by the $s_i$ (recalling (\ref{eq:n0si})). Together with (\ref{eq:dq2}), this gives
\begin{equation}\label{eq:wish}
	dz = \frac1{9\pi q} d y\ .
\end{equation}
Let us see what happens to the positions of D8-stacks in this coordinate. In the massive region, using (\ref{eq:wish}) and (\ref{eq:q2y}) we have
\begin{equation}\label{eq:z1}
	\begin{split}
		\int_{y_{i-1}}^{y_i} dz &= \frac2{3\sqrt{\pi s_l}}\left[\sqrt{y-y_{i-1}+\frac94 \pi\frac {r_{i-1}^2}{s_i}}\right]_{y_{i-1}}^{y_i} \\&=\frac2{3\sqrt{\pi s_l}}\left[ \left(\Delta y_i+\frac94\pi\frac {r_{i-1}^2}{s_i}\right)^{1/2}-\left(\frac94 \pi\frac {r_{i-1}^2}{s_i}\right)^{1/2}\right]=\frac{r_i-r_{i-1}}{s_i}=1\ .
	\end{split}
\end{equation}
In the massless region, $q$ is constant, and $z$ is proportional to $y$; thus it is even simpler to compute, using (\ref{eq:yRL}): 
\begin{equation}\label{eq:zRL}
	z_{\rm R}- z_{\rm L}=N-L-R\ .
\end{equation}

Altogether, (\ref{eq:z1}) and (\ref{eq:zRL}) show that $2 q(z)$ is a piecewise linear function of $z\in[0,N]$, whose graph interpolates the discrete graph of the ranks, just like the solid plot in figure \ref{fig:r-plot}. In other words: \begin{equation}\label{eq:q-lin}
	2q(z)= r_i + s_{i+1} (z-i) \ , \qquad z\in [i,i+1]\ .
\end{equation}
(recall that $s_{i+1}=r_{i+1}-r_i$ and $r_0=r_N\equiv 0$). Now, (\ref{eq:wish}) can be read as $y$ being a primitive of $q(z)$; moreover, from the definition $q\equiv -4y \frac{\sqrt{\beta}}{\beta'}$ we obtain that $\sqrt{\beta}$ is a primitive of $y$:
\begin{equation}\label{eq:q-prim}
	q=\frac1{9\pi} \del_z y \ ,\qquad y=-\frac1{18\pi} \del_z \sqrt{\beta}\ .
\end{equation}
These facts will be important in the holographic match in section \ref{sec:hol}. Integrating (\ref{eq:q-lin}) we find 
\begin{equation}\label{eq:yb-exp}
\begin{split}
	\frac2{9\pi} (y-y_i) &= r_i(z-i) +\frac12 s_{i+1}(z-i)^2 \ ,\\
	-\frac1{(9\pi)^2}\left(\sqrt{\beta}-\sqrt{\beta_i}\right) &=\frac2{9\pi}y_i(z-i)+\frac12 r_i (z-i)^2 +\frac16 s_{i+1}(z-i)^3\ ,
\end{split}
\qquad z\in [i,i+1]\,.
\end{equation}
We determine the integration constants $y_i$  and $\beta_i$ in appendix \ref{app:int}. As a cross-check, notice that in the massless region $s_{i+1}=0$, and $\sqrt{\beta}$ becomes quadratic; this is consistent with (\ref{eq:beta-ml}), recalling that $z$ is proportional to $y$ in the massless region.

Let us also show how the metric looks like in the coordinate $z$ we just introduced:\footnote{The fact that we managed to write the metric in terms of a piecewise linear function is reminiscent of \cite{gaiotto-maldacena}. The ultimate reason is that the combinatorial data are formally the same, but it might be interesting to explore this relationship further.}
\begin{equation}
	\frac1{\pi \sqrt2} ds^2= 8\sqrt{-\frac \alpha{\ddot \alpha}}ds^2_{{\rm AdS}_7}+ \sqrt{-\frac {\ddot \alpha}\alpha} \left(dz^2 + \frac{\alpha^2}{\dot \alpha^2 - 2 \alpha \ddot \alpha} ds^2_{S^2}\right)
	\ ,\qquad \alpha \equiv \sqrt{\beta} \ .
\end{equation}
The dilaton reads
\begin{equation}
	e^\phi=2^{5/4}\pi^{5/2} 3^4 \frac{(-\alpha/\ddot \alpha)^{3/4}}{\sqrt{\dot \alpha^2-2 \alpha \ddot \alpha}}\ .
\end{equation}
Notice that (\ref{eq:q-prim}) implies $\ddot \alpha<0$. We also have
\begin{equation}
	B=\pi \left( -z+\frac{\alpha \dot \alpha}{\dot \alpha^2-2 \alpha \ddot \alpha}\right) {\rm vol}_{S^2}\ ,\qquad F_2 =  \left(\frac{\ddot \alpha}{162 \pi^2}+ \frac{\pi F_0\alpha \dot \alpha}{\dot \alpha^2-2 \alpha \ddot \alpha}\right) {\rm vol}_{S^2}\ .
\end{equation}
The expression for $B$ is now valid both in the massless and massive regions. In the latter we have that $F_2-F_0 B$ is a closed form, as it should be. 


\subsubsection{Holographic limit} 
\label{ssub:hol}

Finally we will identify the conditions under which the solutions of this section have small curvature and string coupling. 
Usually one tends to take large ranks. However, in our case it seems more appropriate to scale the number of \emph{gauge groups}. Intuitively, the idea is that our solutions came from a near-horizon limit of NS5-branes, and the curvature is small when the number $N$ of fivebranes is large. This is even clearer for the massless solution (\ref{eq:beta-ml}), which is a reduction of $N$ M5-branes. 

Indeed one sees from (\ref{eq:y0yN}) that making $N$ very large makes the range of $y$ become large too. This looks promising, but one also sees from (\ref{eq:Deltay}) that the $\Delta y_i$ for $i\le L$ and $i\ge R$  are staying constant. This can be seen even more clearly in the $z$ coordinate introduced in section \ref{ssub:z}: the total range of the $z$ coordinate is $N$, but (\ref{eq:zRL}) shows that only the massless region is expanding; the massive regions stay the same size. In terms of figure \ref{fig:s-plot}, the central region between the two Young diagrams is expanding more and more. A more careful analysis indeed concludes that the D8's are becoming smaller and smaller with respect to the internal volume: the massless region is expanding, pushing the D8's closer and closer to the poles. Thus in this limit we are getting back to the massless solution (\ref{eq:beta-ml}) and the details of the tail of the quiver associated to the massive regions are washed out. 

So we should also rescale the massive regions at the same time as the massless one; in other words we should take  
\begin{equation}\label{eq:holo_limit_gen}
N, L, R\to \infty \quad \mathrm{with} \quad \frac{L}{N},\frac{R}{N} \quad \mathrm{constant}.
\end{equation}
We will refer to this as the \emph{holographic limit} in the following. 

\begin{figure}[ht]
	\centering
		\includegraphics[width=15cm]{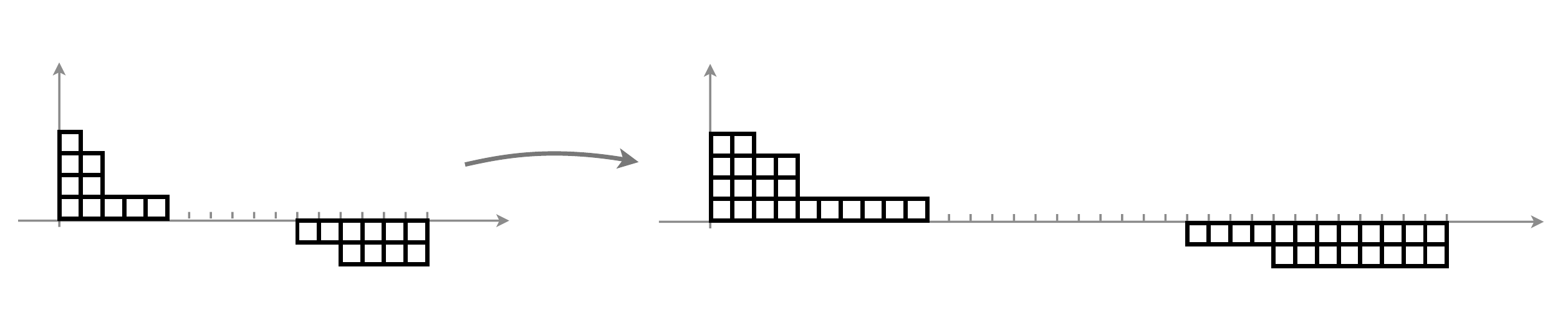}
	\caption{\small The effect of the map (\ref{eq:hor-stretch}) on the plot in figure \ref{fig:s-plot}, for $n=2$.}
	\label{fig:hor-stretch}
\end{figure}

A convenient way to reach this holographic limit is to use a symmetry of the system of BPS equations of supergravity that was pointed out in \cite[Eq.~(4.3)]{gaiotto-t-6d}. In our present language, it reads
\begin{equation}\label{eq:hor-stretch}
	N\to n N \ ,\qquad	 \mu_i \to n \mu_i\ .
\end{equation}
In other words, as well as rescaling $N$, we also rescale the D6-charges of all the D8-brane stacks. We now see in the $z$ coordinate that the positions of the D8-branes, and the size of the massless region, have been simultaneously rescaled by $n$. It may be helpful to visualize this with the help of the action on the $s_i$ plot, shown in figure \ref{fig:hor-stretch} for $n=2$. 

The map (\ref{eq:hor-stretch}) has the effect
\begin{equation}
	e^{2A}\to n e^{2A} \ ,\qquad e^{2 \phi} \to \frac1n e^{2 \phi}\ ,
\end{equation}
and thus can be used to make both curvature and string coupling small. However, as we have just argued,  the D8-branes are rescaled proportionally, and the overall shape of the solution is preserved.

In conclusion, our \emph{holographic rescaling} $n\to\infty$ in \eqref{eq:hor-stretch} will consist in taking  
\begin{equation}\label{eq:hol-limit}
	N \to \infty \ ,\qquad \frac{\mu_i}N = {\rm const}.
\end{equation}
This particular rescaling keeps finite the number of D8-branes, so that in the limit the solution looks for example like the one in figure \ref{fig:M3-D8}. This will be our main focus in what follows. However, it is also possible to consider other limiting procedures, where the solution ends up having infinitely many D8-branes, with a continuous distribution in the rescaled coordinate $z/N$ as $N\to\infty$.  As it turns out, our holographic comparison will also work in such cases, as long as \eqref{eq:holo_limit_gen} is satisfied.

Let us also quickly consider the symmetry \cite[Eq.~(4.2)]{gaiotto-t-6d}. This corresponds to stretching the Young diagrams \emph{vertically}, without stretching them horizontally nor changing the massless region. It rescales all the ranks, $r_i \to n r_i$ (therefore $k \to n k$), and does not change the number of gauge groups. This rescaling achieves $e^{\phi} \to \frac1n e^{\phi}$; thus it can be used to make the string coupling small, but it does not act on the curvature. More generally, large $k$ ensures small string coupling in IIA, but as we will see this is not necessary for our holographic match, as long as $N$ is large.  For this reason, we prefer to use the rescaling (\ref{eq:hol-limit}) to reach the holographic regime.





\section{Anomaly computation in field theory} 
\label{sec:an}

We will now compute the $a$ anomaly of the field theories described in section \ref{sub:ft}. In section \ref{sub:a6}, a straightforward generalization of computations in \cite{intriligator-a6,osty-a6} (with a crucial ingredient from \cite{cordova-dumitrescu-intriligator-a6}) will allow us to isolate the leading term in the holographic limit. In section \ref{sub:lead} we will then focus on how to compute that leading term for concrete Young diagrams. 

\subsection{Anomaly computation} 
\label{sub:a6}

The Weyl anomaly can be expressed in any even dimensions as \cite{deser-schwimmer} $\langle T^\mu_\mu \rangle \sim a E + \sum_i c_i I_i$ up to total derivatives that can be reabsorbed by local counterterms. Here $E$ is the Euler density, and $I_i$ are invariants built out of the Weyl tensor; in six dimensions there are three of them \cite{bonora-bregola-pasti,deser-schwimmer}.
 $a$ has a special role: it does not break scale invariance, and has the ``$a$-theorem'' property of decreasing in an RG flow in two \cite{zamolodchikov} and four \cite{komargodski-schwimmer, komargodski} dimensions. Intuitively, it gives a measure of the ``number of degrees of freedom'' of a CFT. Importantly for us, it can be identified holographically, as we will see in section \ref{sec:hol}. 

The logic that allows to compute $a$ for our class of theories is the following.\footnote{For some theories other methods are available. One can compute the coefficients in (\ref{eq:abcd}) below using anomaly inflow \cite{duff-liu-minasian,Witten:1996hc,freed-harvey-minasian-moore,harvey-minasian-moore,Ohmori:2014pca};        or, in the case of the $(2,0)$ theories, one can use maximal supersymmetry to constrain higher-derivative terms that contribute to $a$ \cite{maxfield-sethi,cordova-dumitrescu-yin}.} First of all, like in four dimensions \cite{anselmi-etal}, one expects it to be related by supersymmetry to the R-symmetry anomaly. The precise formula was actually found only recently \cite{cordova-dumitrescu-intriligator-a6}:%
\footnote{Similar formulas for the three $c_i$ have been recently proposed in \cite{beccaria-tseytlin-ci}. Also, \cite{heckman-rudelius} have used the classification in \cite{heckman-morrison-rudelius-vafa} to give evidence that other combinations of the coefficients in (\ref{eq:abcd}) might be monotonic in RG flows.}
\begin{equation}\label{eq:abcd}
	a= \frac{16}7 (\alpha - \beta + \gamma)+ \frac 67 \delta\ ,
\end{equation} 
where the Greek letters refer to the coefficients in the anomaly polynomial%
\footnote{We omit theory-dependent flavor anomalies, 
since they do not play a role in the following.}
\begin{equation}\label{eq:I8}
	I_8 = \frac1{24}\left(\alpha c_2^2(R)+ \beta c_2(R) p_1 + \gamma p_1^2 + \delta p_2\right) \ .
\end{equation}
With a common abuse of notation we denote by $c_2(R)$, $p_i$ the densities which integrated give the Chern class of the R-symmetry bundle and the Pontryagin classes of the tangent bundle. Thus $\alpha$ is an R-symmetry anomaly, $\gamma$ and $\delta$ are gravitational anomalies, and $\beta$ is mixed. We will see, in any case, that the leading coefficient in (\ref{eq:abcd}) arises from $\alpha$. 
 
An anomaly should not change under RG flow. In general, however, a symmetry might be broken along a flow, and restored only in the IR; or, it might mix with new symmetries that emerge in the IR. However, the effective theories considered in section \ref{sub:ft} are obtained by flowing to the tensor branch, and neither the SU(2) R-symmetry nor diffeomorphisms are broken along the flow. So we know that the anomaly polynomial of the effective theories should in fact be the same as the anomaly polynomial of the CFT in the UV. 

One might be puzzled by this conclusion, given that we described $a$ as a measure of the number of degrees of freedom. When $N$ NS5's coincide one expects a Weyl anomaly scaling with $N^3$ (just like for M5's), while the fields in the effective action are only $\sim N$ in number. However, for these theories the GSWS mechanism that cancels gauge anomalies also gives a large contribution to the anomaly polynomial $I_8$ for global symmetries; it is this contribution that gives the expected $N^3$ behavior. 

Let us see this more concretely, generalizing straightforwardly a computation in \cite{osty-a6}.  
Before taking into account the GSWS terms, the contributions of vector, hyper and tensor multiplets are
\begin{align}\label{eq:Ivht}
	I^{\rm vec}&= -\frac1{24}\sum_{i=1}^{N-1}\left[2 r_i {\rm tr}F_i^4 + 6 ({\rm tr}F_i^2)^2 + 12 r_i c_2 {\rm tr}F_i^2 +(r_i^2-1) c_2^2 +\right.\nonumber\\
	&\left.+\frac{p_1}2 (2 r_i {\rm tr}F_i^2 +(r_i^2-1) c_2) +\frac{r_i^2-1}{240}(7 p_1^2-4 p_2) )\right]\ ,\nonumber\\
	I^{\rm hyp}&=\frac1{24}\sum_{i=1}^{N-2}\left[ 
	r_{i+1} {\rm tr}F_i^4+ r_i {\rm tr}F_{i+1}^4+ 6 {\rm tr}F_i^2{\rm tr}F_{i+1}^2
	+\frac{p_1}2 (r_i {\rm tr}F_{i+1}^2 +r_{i+1} {\rm tr}F_i^2) \right.\\
	&\left.+ \frac{r_i r_{i+1}}{240}(7 p_1^2 -4 p_2)\right]+\frac1{24}\sum_{i=1}^{N-1}\left[f_i {\rm tr}F_i^4+\frac{p_1}2 f_i {\rm tr}F_i^2 +\frac{f_i r_i}{240}(7 p_1^2-4 p_2)\right]\ ,\nonumber\\
	I^{\rm tens}&= \frac1{24}(N-1)\left(c_2^2 +\frac12 c_2 p_1 +\frac1{240}(23 p_1^2-116 p_2) \right)\ , \nonumber
\end{align}
where $c_2\equiv c_2(R)$, $F_i$ is the field-strength of the $i$-th gauge group and ${\rm tr}$ denotes the trace in the fundamental representation. Note that we only included the $N-1$ tensor multiplets for the relative positions of the NS5-branes in the $x^6$ direction, and disregarded the free tensor multiplet for the center of mass motion, which decouples from the CFT. 
The total reads 
\begin{align}\label{eq:Itot}
	I^{\rm tot}&=\frac1{24} \Bigg( \sum_{i=1}^{N-1}\left[ (-2 r_i + r_{i-1}+r_{i+1}+f_i)\left({\rm tr}F_i^4 +\frac{p_1}2 {\rm tr}F_i^2\right)-12 r_i c_2 {\rm tr}F_i^2\right]\nonumber\\
	&-3 \sum_{i,j}C_{ij} {\rm tr}F_i^2{\rm tr}F_j^2 +\left(2(N-1)-\sum_i r_i^2\right)\left(c_2^2 +\frac12 c_2p_1\right) +\frac {N-1}{240} (23 p_1^2-116 p_2)\nonumber\\
	&+\frac{7 p_1^2-4 p_2}{240}\left(N-1+\frac12\sum_i r_i(-2r_i + r_{i-1}+r_{i+1} +2f_i)\right)\Bigg) \ .
\end{align}  
Here
\begin{equation}
	 C_{ij}= 2 \delta_{ij}- \delta_{i,j-1}- \delta_{i,j+1}
\end{equation}
is the Cartan matrix of $A_{N-1}$; 
its appearance will be crucial later on.

The presence of $F_i$ in (\ref{eq:Itot}) indicates that 
we have not yet canceled the gauge anomalies. The terms ${\rm tr}F_i^4$ and $p_1 {\rm tr}F_i^2$ can be canceled quite simply by requiring (\ref{eq:rifi}). 
  
Canceling the terms $C_{ij} {\rm tr}F_i^2{\rm tr}F_j^2$ and $r_i c_2 {\rm tr}F_i^2$ is more challenging. Completing the square, we can rewrite those two terms as 
\begin{equation}\label{eq:comp-sq}
	-\frac18 C_{ij}I_i I_j + \frac12 C^{-1}_{ij}r_i r_j c_2^2\ ,\qquad I_i \equiv {\rm tr}F_i^2  + 2c_2 C^{-1}_{ij}r_j \ 
\end{equation}
where now a sum over repeated indices is understood.
Of these, only the first term contains the gauge field-strength.
Its structure as an inner product strongly suggests that it should be canceled by a GSWS mechanism, as done in \cite{green-schwarz-west,sagnotti} for theories coupled to gravity; as in \cite{intriligator-a6,osty-a6}, we will assume this to be the case. So we assume that the Lagrangian contains a term
\begin{equation}\label{eq:gs}
	{\cal L}_{\rm GS}=\frac18 C_{ij} b_i I_j\ ,
\end{equation}
where the $N-1$ two-form potentials $b_i$ ($i=1,\ldots,N-1$) are related to the $N$ potentials $B_i$ associated to the $N$ NS5-branes according to $B_i-B_{i+1}=C_{ij} b_j$, the change of basis from simple roots to fundamental weights of $A_{N-1}$.%
\footnote{The decoupled center of mass mode is not involved in the GSWS mechanism.}  
The two-form potentials $b_i$ transform under gauge transformations according to $\delta b_i = I^1_{2i}$, where the 2-form $I^1_{2i}$ is related to the 4-form $I_i$ by the descent mechanism:
\begin{equation}
	I_i = d I_{3i} \ ,\qquad \delta I_{3i}= dI^1_{2i}\ .
\end{equation}
Explicitly, $I^1_{2i}= 
{\rm tr}(\lambda_i dA_i)+ {\rm tr}(\lambda^{({\rm R})} dA^{({\rm R})}) C^{-1}_{ij}r_j$, where $A_i$ and $\lambda_i$ are the connections and parameters for the SU$(r_i)$ gauge symmetries, and similarly $A^{({\rm R})}$ and $\lambda^{({\rm R})}$ are a background connection and parameter for the SU$(2)_{\rm R}$ global symmetry, that we included to manifest the SU$(2)_{\rm R}$ anomaly. 

Likewise, $I_8 = d I_7$, $\delta I_7=dI_6^1$; $I_6^1$ is the anomaly we want to cancel. Taking $I_7=-\frac18C_{ij}I_{3i}I_j$, $I^1_6= -\frac18C_{ij}I^1_{2i}I_j$, we see that indeed (\ref{eq:gs}) does the job. Thus, of the two terms in (\ref{eq:comp-sq}), only the second, $\frac12 C^{-1}_{ij}r_i r_j c_2^2$, remains. This term will have a crucial role.

Taking all this into account, we can now go back to (\ref{eq:Itot}) and collect the various terms that have survived in the four coefficients of (\ref{eq:I8}): 
\begin{equation}
\begin{split}
	&\alpha= 12 \sum_{i,j}C^{-1}_{ij}r_i r_j +2(N-1) -\sum_i r_i^2 \ ,\qquad
	\beta = N-1-\frac12 \sum_i r_i^2 \ ,\\
	&\gamma = \frac1{240}\left(\frac72 \sum_i r_i f_i +30 (N-1)\right)\ ,\qquad
		\delta= -\frac1{120}\left(\sum_i r_i f_i + 60 (N-1)\right)\ .
\end{split}
\end{equation}
Notice that $\gamma$ and $\delta$ are equal to those one would have with $N-1$ tensor multiplets and $d_{\rm H}$ hypermultiplets, where $d_{\rm H}= \frac12\sum_i r_i f_i+N-1$ is the dimension of the Higgs moduli space of the quiver theory. This can be explained by the presence of a flow to a mixed Higgs-tensor branch \cite{cordova-dumitrescu-intriligator-a6}.%
\footnote{$d_{\rm H}$ gets naturally assembled in the terms $(7p_1^2-4p_2)$ as $\sum \#(\text{hypers})-\sum{\rm dim}(\text{gauge groups})$. We thank N.~Mekareeya for discussions about this point. See also \cite{zafrir}.} Using now (\ref{eq:abcd}), we get
\begin{equation}\label{eq:a}
	a=\frac{16}7\left(12\sum_{i,j}C^{-1}_{ij}r_i r_j-\frac12 \sum_i r_i^2 +\frac{11}{960}\sum_i r_i f_i +\frac{15}{16}(N-1)\right)\ .
\end{equation}


\subsection{Leading behavior in the holographic limit} 
\label{sub:lead}

In preparation for our comparison with the holographic computation in section \ref{sec:hol}, we will now isolate the leading behavior of $a$ in (\ref{eq:a}) in the holographic limit (\ref{eq:holo_limit_gen}). 

\begin{figure}[ht]
\centering	
\includegraphics[width=10cm]{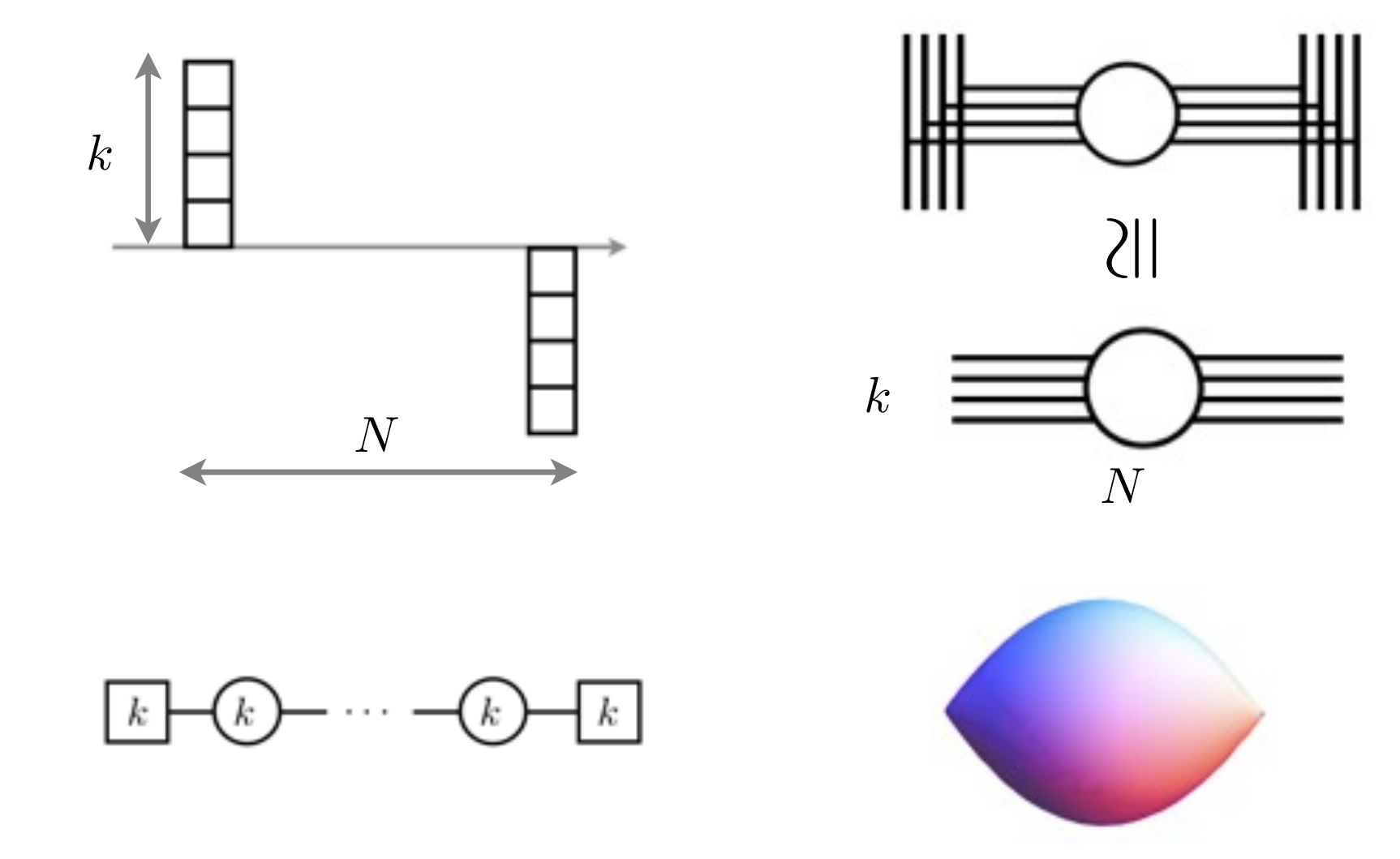}
	\caption{\small A theory that is dual to the massless solution in the holographic limit. From the top left, anticlockwise, we show: the Young diagrams, the quiver, a sketch of the internal space $M_3$, and the brane configuration; cf.~the general case in figures \ref{fig:s-plot}, \ref{fig:quiver}, \ref{fig:M3-D8}, \ref{fig:D8-out}. The brane picture is shown in the version that follows from the general correspondence reviewed in section \ref{sub:ads}, as well as in an alternative version, using the equivalence of a D8-brane with one D6 attached and a semi-infinite D6 \cite{gaiotto-witten-1}. Taking the general correspondence literally, one would see in the gravity solution two D8 stacks with D6-charges $\pm1$, but in the holographic limit these become so small as to be indistinguishable from a D6 and an anti-D6 stack. 
}
	\label{fig:massless}
\end{figure}

In order to do so, we will present a few alternative expressions for the various terms in (\ref{eq:a}). However, we can get some intuition by looking at the case where all ranks are equal, $r_i = k$. 
According to the general correspondence explained in section \ref{sub:ads}, this should correspond to two D8 stacks of charge $\mu=\pm1$, separated by a massless region. As described in section \ref{ssub:hol}, if we keep $\mu$ fixed at $\pm1$ while sending $N\to \infty$, the D8-branes become smaller and smaller,%
\footnote{We will see later what happens when one instead rescales $\mu$ at the same time as $N$.} and the solution is actually well approximated by the massless solution (\ref{eq:beta-ml}), which has a stack of D6-branes at one pole, and a stack of anti-D6-branes at the other; see figure \ref{fig:massless} for a summary of this case. On the field theory side, the computation for this case was already performed in \cite{osty-a6}, where it was pointed out that 
\begin{equation}\label{eq:strange}
	\sum_{i,j} C^{-1}_{ij}= \frac1{12}(N^3-N)\ .
\end{equation}
Thus the leading term in (\ref{eq:a}) is given by $\sum_{i,j}C^{-1}_{ij}r_i r_j\sim \frac1{12}k^2N^3$; the term $\sum_i r_i^2=k^2(N-1)$ grows less fast at large $N$, and the other terms even less so. 

We will now evaluate these terms in general. Let us start from
\begin{equation}\label{eq:Crr}
	\sum_{i,j}C^{-1}_{ij}r_i r_j\ ,
\end{equation}
which will turn out to give the leading contribution, like in the example we just examined. We first need an expression for $C^{-1}$. We obtain
\begin{equation}
	C^{-1}_{ij}= \frac1N \left\{\begin{array}{cc}
		i (N-j)\ ,\qquad &i \le j\ ,\\
		j (N-i)\ ,\qquad &i \ge j\ .
	\end{array} \right.
\end{equation}
Thus (\ref{eq:Crr}) can be written as 
\begin{equation}\label{eq:Crrsum}
	\sum_{i,j} C^{-1}_{ij}r_i r_j=\frac1N \left(\sum_i i (N-i) r_i^2 + 2 \sum_{i<j} i (N-j) r_i r_j \right)\ .
\end{equation}
The large $N$ scaling of \eqref{eq:Crrsum} can be quickly estimated using $i\sim N$ and $\sum_i\sim N$ (since the quiver has length $\sim N$), which implies $C^{-1}_{ij}\sim N$ using \eqref{eq:Crr}. Then \eqref{eq:Crrsum} scales like $N^3$ due to the off-diagonal terms. Similarly, the remaining terms in \eqref{eq:a} are estimated to scale like $N$ in the large $N$ limit.

\begin{figure}[ht]
\centering	
\includegraphics[width=12cm]{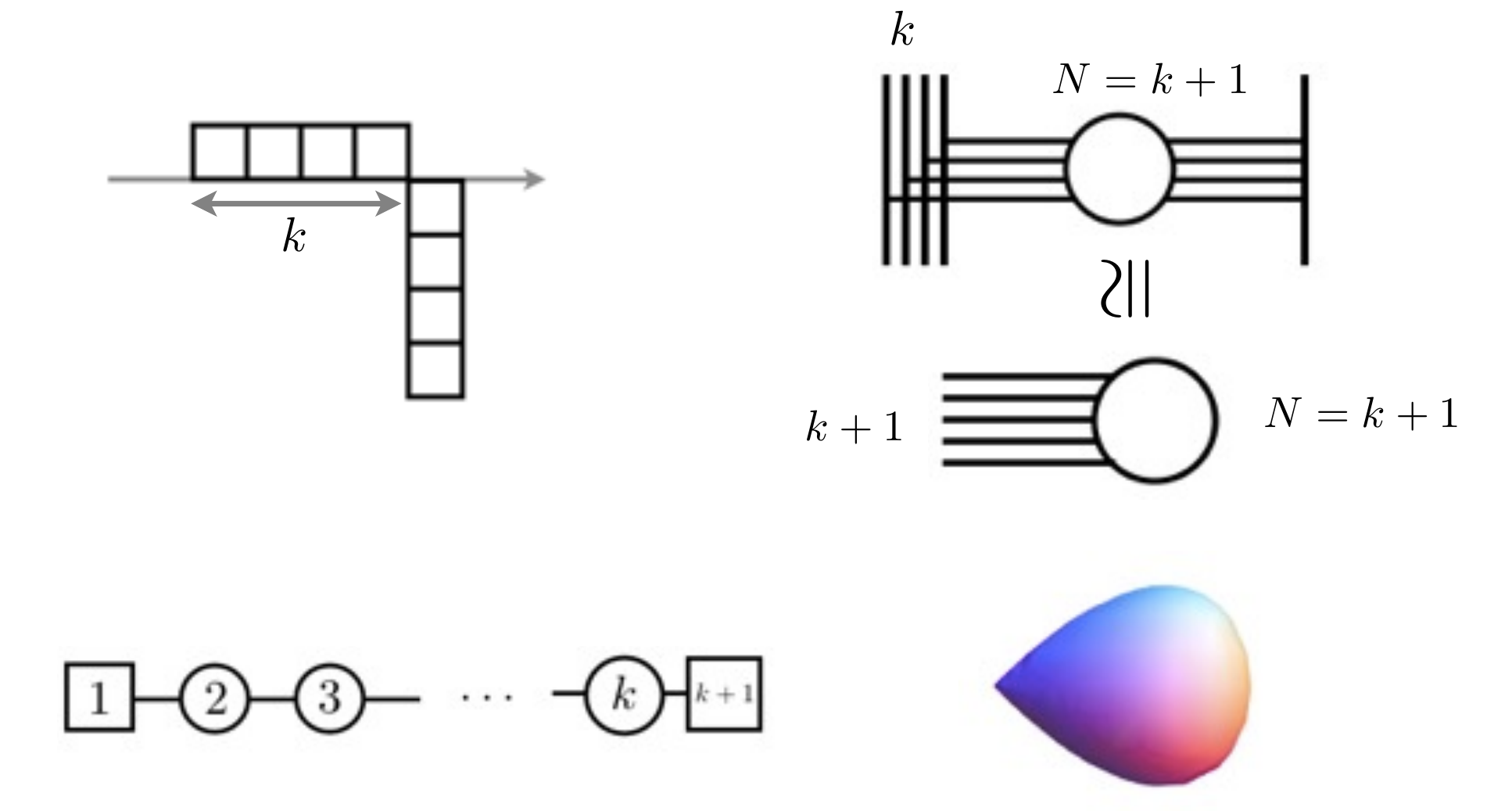}
	\caption{\small A theory dual to the ``simple massive solution'' in \cite{10letter,afrt}.  From the Young diagram picture one sees that there is no massless region. We show the brane picture that follows from the general correspondence, and a simpler one that is obtained by applying Hanany--Witten rules and the equivalence of a D8 with a D6 ending on it with a semi-infinite D6. As in figure \ref{fig:massless}, in the holographic limit the solution is indistinguishable from one with a single D6 stack.
}
	\label{fig:1D6}
\end{figure}

Next we are going to isolate the contribution of the central plateau from those of the two lateral tails. To do so, we can break up each of the sums in (\ref{eq:Crrsum}) in contributions from 1 to $L$, from $L+1$ to $R-1$, and from $R$ to $N-1$. We will describe the result at leading order in $N$, $L$ and $R$, since these are all large in the holographic limit (\ref{eq:holo_limit_gen}): 
\begin{align}\label{eq:Crrlong}
	N \sum C^{-1}_{ij}r_i r_j&\sim \frac{k^2}{12} \left(N-L-R\right)^2\left( N^2+ 2(L+R)N -3 (L-R)^2\right) \nonumber\\
	 &+ \frac k2 (N-L-R) \left((N-L+R)\sum_{i=1}^L i r_i + (N+L-R)\sum_{i=1}^R i r_{N-i} \right)\nonumber\\
	&+ 2\sum_{i=1}^L i r_i \sum_{j=1}^R j r_{N-j}+ 
	\sum_{i=1}^L i(N-i) r_i^2 + 2\sum_{i<j\le L} i(N-j) r_i r_j\\
	&+\sum_{i=1}^R i(N-i) r_{N-i}^2 + 2\sum_{i<j\le R} i(N-j) r_{N-i} r_{N-j}\ .\nonumber	
\end{align}
In the first line of this formula we start seeing a cubic scaling with $N$ for $\sum_{i,j} C^{-1}_{ij} r_i r_j$, generalizing (\ref{eq:strange}). 
The remaining parts of this formula can be estimated to be of the same order, but are still complicated. To obtain a formula that might be useful in particular cases, one possibility is to reexpress everything in terms of the $f_i$. The advantage of doing this is that, while all the $r_i\neq0$, often only a few $f_i$ are non-zero, as the example in figure \ref{fig:quiver} illustrates. This becomes even more true under the holographic rescaling (\ref{eq:hol-limit}): the non-vanishing $f_i$ are associated with the D8-stacks, whose number stays fixed under the rescaling. After a lengthy computation we find 
\begin{subequations}\label{eq:rrleft}
\begin{equation}
	\sum_{i=1}^L i r_i \sim \frac16\sum_{i=1}^L i(3L^2-i^2) f_i \ 
\end{equation}
and 
\begin{align}
	\sum_{i=1}^L &i(N-i) r_i^2 + 2\sum_{i<j\le L} i(N-j) r_i r_j \sim \sum_{i=1}^L\sum_{j=1}^L M_{ij}f_i f_j\ ,\\ 
&M_{ij} \equiv \frac N{120} (40 i j L^3-20 ij (i^2 + j^2 )L + 3(i^5+j^5)-5 i j (i^3+j^3)+10 i^2 j^2 (i+j)) \nonumber \\
	&+ \frac1{360}(-90 i j L^4 +30 i j (i^2+j^2) L^2 - 4 (i^6+j^6)+9 i j (i^4 +j^4) -20i^3 j^3)\ ,
	\nonumber
\end{align}
\end{subequations}
assuming $i$ and $j$ are also being rescaled as $N$, as in (\ref{eq:hol-limit}). Similar formulas hold for the $R\le i \le N-1$ region.

\begin{figure}[ht]
\centering	
\includegraphics[width=14cm]{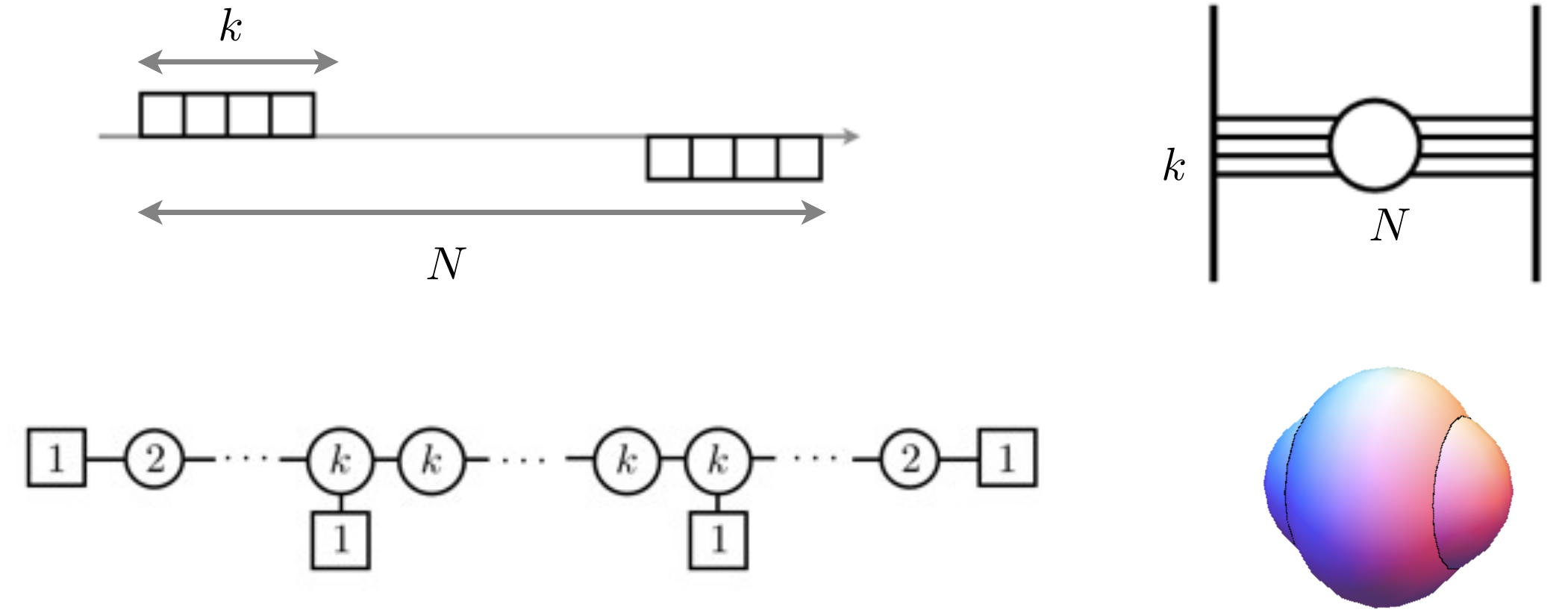}
	\caption{\small The theory dual to two symmetric D8s, of D6 charges $\pm\mu=\pm k$. In this case we have taken both $\mu$ and $N$ to be large and of the same order, just as prescribed in (\ref{eq:hol-limit}).}
	\label{fig:2D8}
\end{figure}

We can evaluate the remaining terms in (\ref{eq:a}) using a similar strategy; it becomes immediately clear that they are subleading. For example, at leading order  $\sum_i r_i^2 \sim (N-L-R) k^2 + \sum_{i=1}^L \sum_{j=1}^L m^L_{ij} f_i f_j + \sum_{i=1}^L \sum_{j=1}^L m^R_{ij} f_{N-i}f_{N-j}$, where $m^L_{ij}=-\frac1{12}(i+j)^3 + i j L$ and similarly for $m^R$. This is subleading with respect to (\ref{eq:Crrlong}), (\ref{eq:rrleft}). So in fact
\begin{equation}\label{eq:aC}
	a \sim \frac{192}7 \sum C^{-1}_{ij}r_i r_j\ .
\end{equation}

The $c_i$ coefficients of the Weyl anomaly can be similarly computed using their linear relations to the coefficients of R-symmetry and diffeomorphism anomalies \cite{beccaria-tseytlin-ci}. Since the coefficients $\beta$, $\gamma$ and $\delta$ in the anomaly polynomial \eqref{eq:I8} are subleading to $\alpha$, in the holographic limit the $c_i$ Weyl anomaly coefficients are all proportional to $a$. 
Specifically we get $c_1\sim -\frac 7{12} a$, $c_2\sim \frac14 c_1$, $c_3\sim-\frac1{12}c_1$. Notice that the ratios between the $c_i$ are the same as the ones for the $(2,0)$ theory \cite{bastianelli-frolov-tseytlin}. 

While these formulas are still very complicated in the most general case, they do become relatively simple in particular examples. Let us apply it to two cases which have already been considered in \cite{10letter}. The first is shown in figure \ref{fig:1D6}. In this case the general formula gives $a \sim \frac{16}7 \cdot\frac4{15} k^5$. The gravity computation in \cite{10letter} was a bit different from the one giving $a$, but it is proportional to it, as we will review in section \ref{sec:hol}. If one normalizes the result against the massless theory we considered around (\ref{eq:strange}), we see that our current result exactly matches the one in \cite{10letter}. Another case is when we have two symmetric stacks of $n_0=k/\mu$ D8-branes of D6-charges $\pm \mu$, surrounding a massless region of D6-charge $k$; see figure \ref{fig:2D8} for the case $k=\mu$. In this case (\ref{eq:Crrlong})--(\ref{eq:aC}) give
\begin{equation}\label{eq:a-ex2}
	a \sim \frac{16}7 k^2\left(N^3-4 N \mu^2 + \frac{16}5 \mu^3\right)\ .
\end{equation}
Again, and more strikingly, this precisely agrees with \cite[Eq.(21)]{10letter}. (Recall also that $\mu\sim N$, as in our comment after (\ref{eq:a-ex}) corresponding to the case $\mu=k$.)




\section{Holographic match} 
\label{sec:hol}

In this section we will compute $a$ from the gravity solutions reviewed in section \ref{sub:ads}, and compare them with the results of section \ref{sec:an} in the holographic limit.

\subsection{Holographic anomaly computation} 
\label{sub:ahol}

The computation of $a$ from gravity was first described in \cite{henningson-skenderis} in various dimensions, after an idea in \cite{witten-hol}. In six dimensions, the computation is directly relevant for the $(2,0)$ A$_N$ or D$_N$ theory, but it is in fact very general and can easily be adapted to our needs. 

Here is a quick review of the computation. The starting point is the seven-dimensional Einstein action $\frac1{16\pi G_{\rm N}}\int d^7x \sqrt{g_7} (R_7 + \Lambda) +$ boundary terms. The metric is written as $ds^2= \frac{l^2}{r^2} (dr^2 + r^2 g^{(6)}_{ij}dx^i dx^j+ \ldots)$; $g^{(6)}_{ij}dx^i dx^j$ is the metric on the boundary. The $\ldots$ are terms that go to zero at the boundary $r=0$, which can be determined in terms of $g^{(6)}_{ij}$ by the equations of motion. The presence of a $\log(r)$ in one of these terms generates the Weyl anomaly, which in the end is of the form $\langle T^\mu_\mu \rangle = \frac{l^5}{G_{\rm N}}\times$ a polynomial in the Riemann tensor of $g^{(6)}_{ij}$ and its derivatives.\footnote{This dependence on $l$ gives another argument that $a$ should decrease in RG flows \cite{girardello-petrini-porrati-zaffaroni,freedman-gubser-pilch-warner}, one that should also hold in  six dimensions, although efforts to prove this directly in field theory have so far been inconclusive \cite{elvang-etal}.}

Now, as also remarked in \cite{gubser-einstein} for the four-dimensional case, in this computation the details of the gravity solution enter only through Newton's constant $G_{\rm N}$. This would be simply proportional to the inverse of the internal volume ${\rm Vol}^{\rm E}(M_3)$ in Einstein frame (the frame used in \cite{henningson-skenderis}). Thus the relevant quantity would be
\begin{equation}
	l^5{\rm Vol}^{\rm E}(M_3)\ ,
\end{equation}
where $l$ is the AdS$_7$ radius. The solutions in section \ref{sub:ads}, however, are warped products: as one can see from (\ref{eq:ds2}), the AdS$_7$ radius is in fact the warping function $e^A$, which depends on the coordinates of $M_3$. In this situation, $l^5$ should be read as the average of $e^{5A}$ over $M_3$. Finally, we should translate our results in the string frame (which we used in section \ref{sub:ads}), recalling $g^{\rm E}_{MN} = e^{-\phi/2}g^{\rm str}_{MN}$; that gives $5+3$ powers of $e^{-\phi/4}$.  This leads us to taking the average of $e^{5A-2 \phi}$ over the internal manifold. This integral indeed scales with the expected $k^2 N^3$ in the case of the massless solution; the $k=1$ case is simply the reduction of AdS$_7\times S^4$, and one can use this case \cite{henningson-skenderis,bastianelli-frolov-tseytlin} to fix the overall factor. 
All in all $a$ reads
\begin{equation}\label{eq:F}
	a_{\rm hol}=\frac3{56\pi^4}\int_{M_3} e^{5A-2\phi}{\rm vol}_3\ .
\end{equation}

The same integral (up to an overall factor) already appeared in \cite{gaiotto-t-6d,10letter} with a slightly different interpretation, namely as the coefficient ${\cal F}_0$ in the free energy ${\cal F}= {\cal F}_0 {\rm Vol} T^6$. This is an alternative measure of the number of degrees of freedom: although it has the advantage of also being defined in odd dimensions, it is perhaps not surprising that in even dimension it is proportional to the Weyl anomaly $a$.

Let us stress once again that (\ref{eq:F}) is only the supergravity contribution, without string theory corrections. For example, in the case of AdS$_7\times S^4$, it gives the leading order $a\sim \frac{16}7 N^3$ \cite{henningson-skenderis,bastianelli-frolov-tseytlin} (again in the convention where $a=1$ for a single $(2,0)$ tensor). The full result is in fact $a=\frac17(16 N^3-9N)$ (which indeed gives $a=1$ for $N=1$); the $-\frac97 N$ term comes from higher curvature corrections  \cite{tseytlin-R4}. This linear term in $N$ is subleading and is not to be confused with the term linear in $N$ in expressions such as (\ref{eq:a-ex2}), which is multiplied by a further large $\mu^2$, and which originates from (\ref{eq:F}) \cite{10letter}. 


In the next subsections, we will evaluate (\ref{eq:F}) for the solutions in \ref{sub:ads}, and compare it with the results in section \ref{sec:an}.


\subsection{The match as a continuum limit} 
\label{sub:quick}

In this section we will give a first argument showing why the internal volume (\ref{eq:F}) agrees with the $a$ anomaly (\ref{eq:a}) in the large $N$ limit.
In section \ref{sub:det} we will present a more detailed comparison. 

Recall that we concluded in section \ref{sub:lead} that $a$  is proportional at leading order to $\sum_{i,j}C^{-1}_{ij} r_i r_j$. Up to a small discrepancy in the very last column, we can write $C\sim-\del \del^*$, where $\del$ and $\del^*$ are discrete derivative operators defined after (\ref{eq:rifi}). In other words, $C$ is a discrete second derivative. So schematically we can write 
\begin{equation}\label{eq:adel2}
	a \sim -\frac{192}7 \sum_i r_i\left(\frac1{\del\del^*}r\right)_i\ .
\end{equation}

Now let us turn to the gravity computation (\ref{eq:F}). Using (\ref{eq:ds2}) and (\ref{eq:ephi}), we evaluate
\begin{equation}
	a_{\rm hol} = \frac{128}{7\cdot 3^5 \pi^3}\int \sqrt{\beta} dy\ .
\end{equation}
This can also be rewritten in the $z$ coordinate using (\ref{eq:q-prim}): 
\begin{equation}\label{eq:ag}
	a_{\rm hol}= \frac{128}{189 \pi^2}\int \sqrt{\beta}\, q\, d z \ .
\end{equation}
Moreover, (\ref{eq:q-prim}) also allows us to write $\sqrt \beta$ as a second primitive of $q$, $\frac1{\del_z^2} q$, so that 
\begin{equation}\label{eq:agdel2}
	a_{\rm hol}= - \frac{192}{7}\int 2q \left(\frac1{\del_z^2} 2q\right) dz\ .
\end{equation}
(To be precise, $\sqrt \beta(z)$ is the second primitive of $q$ that vanishes at the boundary of the interval: $\sqrt\beta\big|_{z=0,N}=0$.) But we saw in section \ref{ssub:z} (see for example (\ref{eq:qiri})) that $2q(z)$ is a piecewise linear function that interpolates the discrete function $r_i$, as in figure \ref{fig:r-plot}. Hence one sees that (\ref{eq:adel2}) should become (\ref{eq:agdel2}) in the $N\to \infty$ limit. 

This schematic argument can be made more precise using the explicit expression for the inverse Cartan matrix. In the large $N$ limit, the leading term in the $a$ Weyl anomaly \eqref{eq:aC} is given by the double sum \eqref{eq:Crrsum}, namely
\begin{equation}\label{match_0}
a \sim \frac{192}{7} \frac1N \left(\sum_i i (N-i) r_i^2 + 2 \sum_{i<j} i (N-j) r_i r_j \right)\ . 
\end{equation}
To extract the leading order as $N\to \infty$, we can take a continuum limit: we replace the position in the linear quiver (normalized by $N$) by a continuous variable, $i/N \leadsto x \in[0,1]$, the numbers of colors by a continuous non-negative concave function, $r_i \leadsto r(x)$, and sums by integrals. In this continuum limit \eqref{match_0} becomes
\begin{equation}\label{match_1}
a \sim \frac{384}{7} N^3 \int_0^1 dy \int_0^y dx~ x(1-y) r(x)r(y) \ . 
\end{equation}
Integrating repeatedly by parts, this double integral can be recast as 
\begin{equation}\label{match_2}
\begin{split}
a &\sim \frac{192}{7} N^3  \left[ \int_0^1 dx~ r^{(-1)}(x)^2 - \left( \int_0^1 dx~ r^{(-1)}(x) \right)^2 \right] \\
&= \frac{192}{7} N^3  \left[ -\int_0^1 dx~ r(x) r^{(-2)}(x) + r^{(-1)}(x)r^{(-2)}(x)\Big|_0^1 - \left(r^{(-2)}(1)-r^{(-2)}(0) \right)^2 \right]
\end{split}
\end{equation}
where $r^{(-n)}(x)$ denotes an $n$-th primitive of $r(x)$. (The result is independent of integration constants, as the first expression involving the variance of $r^{(-1)}(x)$ shows.) If we fix the two integration constants so that $r^{(-2)}(0)=r^{(-2)}(1)=0$, \eqref{match_2} reduces to 
\begin{equation}\label{match_3}
a \sim -\frac{192}{7} N^3 \int_0^1 dx~ r(x) r^{(-2)}(x) \ .
\end{equation}

This formula precisely matches the holographic result \eqref{eq:ag}, using $z=Nx$ and $2q(z)=r(x)$ (recall \eqref{eq:qiri}). We also used (\ref{eq:q-prim}) supplemented with the boundary conditions $\sqrt\beta\big|_{z=0,N}=0$, that are obeyed by the massive IIA solutions and correspond to $r^{(-2)}\big|_{x=0,1}=0$ above.

This argument for the holographic match applies not only in the rescaling limit \eqref{eq:hol-limit}, which leads to a piecewise linear concave function $r(x)$, but also in the more general holographic limit \eqref{eq:holo_limit_gen}. This also allows the presence of infinitely many D8-branes; in this case, using the coordinate $x=z/N$ for $N\to\infty$, the piecewise linear function becomes a general concave function $r(x)$ vanishing at the endpoints $x=0,1$.


\subsection{Detailed comparison} 
\label{sub:det}

Setting our previous argument aside, we will now present the complete computation of (\ref{eq:F}), even before taking the holographic limit. We will then check that the result matches with the field theory prediction (\ref{eq:a}) in the holographic limit. 

We compute the integral (\ref{eq:F}) using the $z$ coordinate expression in (\ref{eq:ag}). We divide the integral in (\ref{eq:F}) in several pieces, between each D8 stack and the next one. In the left massive region, we can compute the contribution from the $(l-1)$-th and $l$-th D8 stack using (\ref{eq:yb-exp}) and (\ref{eq:betai}):
\begin{align}
	 &\frac{128}{189 \pi^2}\int_{y_{l-1}}^{y_l} \sqrt{\beta}\, q\, d z = 
	-\frac{16}7 \left[ \frac4{9 \pi} (r_{l-1}+ 2 r_l + 3(l-1)(r_l+r_{l-1})) \right.\\
	&\left.+\frac15(2 r_l^2 +21 r_l r_{l-1} +12 r_{l-1}^2) + \sum_{i=1}^{l-2}r_i\left(2 r_{l-1}+4r_l+6(l-i-1)(r_l + r_{l-1})\right)\right]~.\nonumber
\end{align}
Summing up all the contributions from the left massive region we get
\begin{align}\label{eq:Fl}
	 &\frac{128}{189 \pi^2}\int_{y_0}^{y_L}  \sqrt{\beta}\, q\, d z = 
	-\frac{32}{35} \left[ k^2 +7 \sum_{l=1}^{L-1}r_l^2+\frac{21}2 k r_{L-1} +5k\sum_{l=1}^{L-2}(3(L-l)-1)r_l\right.\\
	&\left. +\frac{21}2 \sum_{l=1}^{L-1}r_l r_{l-1}+30\sum_{l=1}^{L-1}\sum_{i=1}^{l-2}(l-i)r_l r_i+20\sum_{l=1}^{L-1}r_l r_{l-1} +\frac{10}{9\pi} \left(k(3L-1)+6\sum_{l=1}^{L-1}l r_l\right)\right]\ .\nonumber
\end{align}
The contribution from the right massive region can be obtained from this by replacing $L\to R$, $r_i\to r_{N-i}$, $y_0\to y_{N}$. Both $y_0$ and $y_{N}$ can be found in (\ref{eq:y0yN}).

The contribution from the massless region can be computed by recalling (\ref{eq:beta-ml}). With some manipulations one can write
\begin{equation}\label{eq:Fml}
\begin{split}
	\frac{128}{189 \pi^2} &\int_{y_L}^{y_R} \sqrt{\beta} \, q\, d z = \frac{256}{7 \cdot 3^5\pi^3}
		\left[R_0^6(y_{\rm R}-y_{\rm L})-\frac13(y_{\rm R}^3-y_{\rm L}^3)\right]
		\\
		&=\frac{256}{7\cdot 3^6\pi^3}(y_{\rm R}-y_{\rm L})\left[ \frac32k(\sqrt{\beta_{\rm L}}+\sqrt{\beta_{\rm R}})+(y_{\rm R}-y_{\rm L})^2\right]\ .
\end{split}
\end{equation}	
$y_{\rm R}-y_{\rm L}$ can be found in (\ref{eq:yRL}); $\beta_L$ can be found in (\ref{eq:betaL}), and $\beta_R$ can be found again by $L\to R$, $r_i\to r_{N-i}$, $y_0\to y_{N}$.

We now have to put together the contribution (\ref{eq:Fl}) from the left massive region, the analogue contribution from the right massive region, and the contribution from the central massless region (\ref{eq:Fml}). It is then tedious but straightforward to check that the total sum reduces, in the holographic limit \eqref{eq:holo_limit_gen}, to the field theory result (\ref{eq:Crrlong}). 

\bigskip

This concludes our detailed check of the match between the field theory computation (\ref{eq:abcd}) and the holographic computation (\ref{eq:F}). It confirms our argument of section \ref{sub:quick}. The match provides a strong confirmation of the holographic duality proposed in \cite{gaiotto-t-6d} and reviewed in section \ref{sec:rev}, between the six-dimensional linear quiver theories and the  ``crescent rolls'' AdS$_7$ solutions of \cite{afrt,10letter}.



\section*{Acknowledgements}
We would like to thank B.~Assel, N.~Bobev, C.~Cordova, T.~Dumitrescu, J.~Heckman, C.~Herzog, N.~Lambert, N.~Mekareeya and C.~Vergu for interesting discussions, and A.~Rota for corrections on the manuscript. We thank the Galileo Galilei Institute for Theoretical Physics for hospitality and INFN for partial support for the workshop ``Holographic Methods for Strongly Coupled Systems'', during which this project was conceived. A.T.~is supported in part by INFN, by the MIUR-FIRB grant RBFR10QS5J ``String Theory and Fundamental Interactions'', and by the European Research Council under the European Union's Seventh Framework Program (FP/2007-2013) -- ERC Grant Agreement n. 307286 (XD-STRING).

\appendix

\section{Integration constants} 
\label{app:int}

We will determine here the precise expressions for the integration constants $y_i$ and $\beta_i$ appearing in (\ref{eq:yb-exp}).

We already know quite a bit about the $y_i$: we have determined their differences in (\ref{eq:Deltay}), (\ref{eq:yRL}). 
We have
\begin{equation}\label{eq:yiy0}
	y_i = \left\{
	\begin{array}{ll}
		y_0+\frac94\pi \left(r_i + 2\sum_{j=1}^{i-1}r_j\right) \ ,\qquad &i \le L\ . \\
		y_{N}-\frac94\pi \left( r_{N-i} + 2\sum_{j=1}^{i-1}r_{N-j}\right) \ ,\qquad &i \ge R\ . 
	\end{array}
	\right. 
\end{equation}
So all is left is to determine $y_0$ or $y_{N}$. We have not used (\ref{eq:yRL}) yet, but that is a single equation and it cannot determine both.

In the \emph{symmetric case}, where the Young diagrams $\rho_{\rm L}$ and $\rho_{\rm R}$ are equal, we know that $L=R$, and that $y_{\rm L}= -y_{\rm R}$. This gives an extra equation, and we obtain
\begin{equation}
 y_0 = - y_{N}= \frac94\pi \left(k(2L-N-1) -2\sum_{i=1}^{L-1}r_i \right)\ .\hspace{1cm}
	 \text{(Symmetric models.)}
\end{equation}

In the general (asymmetric) case, things are slightly more complicated. We have made sure that $\beta$ is continuous up to $y_{\rm L}$ starting from the left and up to $y_{\rm R}$ starting from the right, but we should impose that these two values agree upon evolution through the massless region. In order to do so, we go back to (\ref{eq:yb-exp}) and evaluate the expression for $\beta$ at $y=y_{i+1}$; for $i\le L$, for example, we obtain 
\begin{equation}\label{eq:betai}
	-\frac1{(9\pi)^2}\sqrt{\beta_i}=
	\frac2{9\pi} \,i\,y_0+\frac16 r_i + \sum_{j=1}^{i-1} j r_{i-j}\ .
\end{equation}
From this one also gets 
\begin{equation}\label{eq:betaL}
	\sqrt{\beta_{\rm L}}=-\frac{27}2\pi^2 \left[k+\frac{12}{9\pi} L y_0 +6\sum_{i=1}^{L-1}(L-i)r_i  \right] \ .
\end{equation}
In a similar way one gets an expression for $\beta_{\rm R}$. These values have to agree with what one derives from the massless expression (\ref{eq:beta-ml}), namely $\sqrt{\beta_{\rm R}}-\sqrt{\beta_{\rm L}}= \frac2k(y_{\rm L}^2-y_{\rm R}^2)=-9\pi(N-L-R)(y_{\rm R}+y_{\rm L})$. This gives the desired extra equation. In the end one gets
\begin{equation}\label{eq:y0yN}
\begin{split}
	\frac4{9\pi}y_0 = \frac kN (L-N-R)(N+1-L-R) -2\sum_{j=1}^{L-1}r_j+\frac 2N\left(\sum_{j=1}^{L-1}j r_j -\sum_{j=1}^{R-1}j r_{N-j} \right)\ ,\\
	\frac4{9\pi}y_{N} = \frac kN (L+N-R)(N+1-L-R) +2\sum_{j=1}^{L-1}r_{N-j}+\frac 2N\left(\sum_{j=1}^{L-1}j r_j -\sum_{j=1}^{R-1}j r_{N-j} \right)\ .
\end{split}
\end{equation}


\bibliography{at}
\bibliographystyle{at}

\end{document}